\newcommand{\bB}{\mathbf{B}}
\newcommand{\bv}{\mathbf{v}}
\newcommand{\bA}{\mathbf{A}}
\newcommand{\bE}{\mathbf{E}}
\newcommand{\bJ}{\mathbf{J}}
\newcommand{\nn}{\nonumber}

\newcommand{\nb}{\nabla}

\newcommand{\Ac}{ {\mathcal{A}} }

       \newcommand{\Cc}{ {\mathcal{C}} }

       \newcommand{\Fc}{ {\mathcal{F}} }
       \newcommand{\Gc}{ {\mathcal{G}} }
       \newcommand{\Hc}{ {\mathcal{H}} }
       
       \newcommand{\Jc}{ {\mathcal{J}} }
       
       \newcommand{\Lc}{ {\mathcal{L}} }
       \newcommand{\Mc}{ {\mathcal{M}} }
       \newcommand{\Nc}{ {\mathcal{N}} }

       \newcommand{\Qc}{ {\mathcal{Q}} }
       
       \newcommand{\Sc}{ {\mathcal{S}} }
       
       \newcommand{\Uc}{ {\mathcal{U}} }
       
       \newcommand{\Wc}{ {\mathcal{W}} }

       \newcommand{\R}{ {\mathbb{R}} }

       \newcommand{\bq}{\mathbf{q}}
       \newcommand{\bpi}{\boldsymbol{\pi}}
        
       \newcommand{\ba}{\mathbf{a}}
       \newcommand{\bg}{\mathbf{g}}
       
       \newcommand{\bzt}{\boldsymbol{\zeta}}
       \newcommand{\bht}{\boldsymbol{\eta}}
       \newcommand{\bom}{\boldsymbol{\omega}}
       \newcommand{\bw}{\boldsymbol{w}}
       \newcommand{\bx}{\mathbf{x}}
       \newcommand{\bQ}{\mathbf{Q}}
       \newcommand{\bD}{\mathbf{D}}
      \newcommand{\am}{\mathrm{a}}
      \newcommand{\tbzt}{\tilde{\bzt}}
      \newcommand{\tbht}{\tilde{\bht}}
    \newcommand{\bdd}{\boldsymbol{:}}

\documentclass[twocolumn,amsmath,amssymb,aps]{revtex4-1}

\usepackage[utf8]{inputenc}
\usepackage[T1]{fontenc}
\usepackage{amsmath}

\usepackage{mathtools}
\usepackage{amsfonts}
\usepackage{graphicx}% Include figure files
\usepackage{dcolumn}% Align table columns on decimal point
\usepackage{bm}% bold math
\usepackage[export]{adjustbox}
\usepackage{wrapfig}
\usepackage{lipsum}
\usepackage[titletoc,toc,title]{appendix}
\allowdisplaybreaks
\usepackage[colorlinks]{hyperref}
\usepackage{color}

\begin{document}

\title{Energy-Casimir, dynamically accessible,  and Lagrangian stability of extended magnetohydrodynamic equilibria}% Force line breaks with \\

\author{D. A. Kaltsas}
 %\altaffiliation[Also at ]{Physics Department, University of Ioannina}%Lines break automatically or can be forced with \\
 \email{dkaltsas@cc.uoi.gr}
\author{G. N. Throumoulopoulos}%
 \email{gthroum@uoi.gr}
\affiliation{
Department of Physics, University of Ioannina,\\ GR 451 10 Ioannina, Greece
}%
 \author{P. J. Morrison}
 \email{morrison@physics.utexas.edu}
\affiliation{%
Department of Physics and Institute for Fusion Studies,\\ University of Texas, Austin, Texas 78712, USA 
}


\begin{abstract}
The formal stability analysis of Eulerian extended magnetohydrodynamics (XMHD) equilibria is considered within the noncanonical Hamiltonian framework by means of the energy-Casimir variational principle and the dynamically accessible stability method. Specifically, we find explicit sufficient stability conditions for axisymmetric XMHD and Hall MHD (HMHD) equilibria with toroidal flow and for equilibria with arbitrary flows under constrained perturbations. The dynamically accessible, second-order variation of the Hamiltonian, that can potentially provide explicit stability criteria for generic equilibria is also obtained. Moreover, we examine the Lagrangian stability of the general quasi-neutral two-fluid model written in terms of MHD-like variables, by finding the action and the Hamiltonian functionals of the linearized dynamics, working within a mixed Lagrangian-Eulerian framework. Upon neglecting electron mass we derive a HMHD energy principle and in addition, the perturbed induction equation arises from  Hamilton's equations of motion in view of a consistency condition for the relation between the perturbed magnetic potential and the canonical variables.
%\end{description}
\end{abstract}

\pacs{Valid PACS appear here}% PACS, the Physics and Astronomy
\maketitle

%\tableofcontents

%%%%%%%%%%%%%%%%%%%%%%%%%%%%%
%%%%%%%%%%%%%%%%%%%%%%%%%%%%%
%%%%%%%%%%%%%%%%%%%%%%%%%%%%%
\section{Introduction}

The stability of plasma equilibria is crucial for the attainment of long lived states of magnetically confined plasmas, with sufficient confinement of thermal energy for the self-sustained operation of thermonuclear reactors. In general, the most drastic way to lose the confinement of plasma energy is the development of either macroinstabilities,  e.g.,  the current driven kink and the pressure driven ballooning instabilities, associated with plasma disruption (which effectively put upper limits on the attainable pressure and current), or microinstabilities that  result in enhanced turbulence and anomalous transport. Stability analyses are usually performed using the standard MHD energy principle \cite{Bernstein1958} that was generalized  for flowing equilibria in \cite{Frieman1960}.  The stability of stationary plasma states with macroscopic sheared flows, albeit a tough problem from the mathematical point of view,  is  important since it is believed that plasma rotation, either being self-generated or driven externally, may have beneficial effects in terms of confinement.  Indeed plasma flows  are associated with the suppression of turbulence \cite{Terry2000} and the L-H transitions \cite{Wagner2007} observed in Tokamaks. Also, there are many studies proposing that plasma sheared rotation variously affects  the stability properties of Tokamak equilibria in several cases, either inducing stabilization or destabilization (e.g. \ \cite{Wahlberg2000,Miller1995,Chu1998,Chapman2011,Brunetti2017}), with the main destabilizing mechanism being the Kelvin-Helmholtz instability \cite{pjmC91}.\\ \indent
Furthermore, many astrophysical phenomena, such as the development of turbulence in various stages of the solar wind and in magnetized accretion disks, are consequences of flow-driven instabilities such as the Kelvin-Helmholtz (e.g.\ see \cite{Mishin2016}) and the magnetorotational instability (MRI) \cite{Balbus1991}. It is evident that plasma instability is the reason for the emergence of new structures, but, most importantly,  for fusion physics, instabilities  are the main mechanisms behind the undesirable interchange of energy, which should  be sufficiently reduced in fusion experiments. This pursuit is the main reason for performing  stability studies for over sixty years, trying to refine the resulting stability or instability criteria and incorporate as much physics as possible.  %As mentioned above, the main framework within which the majority of equilibrium and stability studies are performed, is  single fluid ideal magnetohydrodynamics (MHD), since it has been  a successful model for the description of macroscopic motions in plasmas. 
\\ \indent
It  is widely agreed that ordinary MHD, despite being a successful model for describing macroscopic phenomena,  provides a rather rough description of plasmas because it neglects the presence of multi-fluid effects. This is especially true  when there exist characteristic length scales comparable to the ion and electron skin depths, e.g.,  due to the presence of current sheets or thin boundary layers.  In such cases multi-fluid models are needed to describe phenomena arising due to the coexistence of different particle species and the decoupling of their respective motions, even at the macroscopic level. Regarding stability, when mode frequencies comparable to the particle gyro-frequencies are present, then MHD becomes clearly an insufficient framework. This intuitive reasoning about the insufficiency of the MHD model is corroborated when  MHD theory fails to predict adequately the experimental observations:  the observed stability of elongated Field Reversed Configurations (FRC) \cite{Ishida1988,Barnes2002} and the high magnetic reconnection rates (see e.g.\ \cite{Birn2001,Andres2016}), are examples where two-fluid models work  significantly better than MHD. Moreover, there exist recent views on Tokamak physics, suggesting that the Hall drift term cannot be neglected both in equilibrium and dynamics computations; also,  it has been suggested that Hall effects may be associated with the pressure pedestals, formed in the L-H transitions \cite{Gourdain2017,Guzdar2005}.\\ \indent
For the reasons described above, very often, we need to invoke multifluid descriptions since they capture finer dynamical effects, taking place in shorter length and temporal scales. %Regarding stability analysis of flowing plasmas though, a two-fluid treatment is an even tougher problem.
 If rotation is neglected the two-fluid effects are incorporated more easily through the multifluid pressure (e.g. see \cite{Hameiri2004}) because no decoupling of electron and ion motion occurs. However, as was stressed earlier, plasma flows are consequential, and therefore, it is important to take them into account. A characteristic consequence of including flows in stability methods based on energy functionals is the nonseparability of the kinetic and potential energy contributions,  rendering the resulting stability criteria sufficient but not necessary. A typical example is the MHD energy principle, which for static equilibria provides a necessary and sufficient condition \cite{Bernstein1958},  while for stationary states \cite{Frieman1960} it provides only sufficient conditions. These are, respectively,  the Lagrange and Dirichlet conditions of Hamiltonian dynamics, as pointed out in \cite{Morrison1998}. As we shall see later, the non-separability is even stiffer in the two-fluid case. Hence, we understand that forming sufficient and necessary stability criteria for flowing equilibria would require the introduction of several restrictions on the equilibrium states or/and the perturbations under consideration. \\ \indent
Given the historical precedent,  it would appear desirable to apply formal stability analysis methods, similar to those originating from the MHD energy principle to flowing multifluid plasma equilibria, because this framework is already well known from  MHD theory and also because this would facilitate  comparisons with  the MHD results. By formal stability we mean an analysis based on a quantity, a kind of energy,  which is conserved by the full nonlinear dynamics of the system.  The first variation of the quantity must vanish and the second variation must be  positive (or negative) definite at equilibrium.  When this is the case,  the second variation serves as a Lyapunov functional for the linear dynamics. %Formal stability is important because it implies  linearized and spectral stability and is a step towards nonlinear stability, which involve questions of existence and convexity \cite{Holm1985,Morrison1998}. 
 At present,  only a limited number of studies have led to appropriate Lyapunov functionals and ultimately to stability conclusions within the two-fluid context, primarily in the Hall MHD (HMHD) limit \cite{Holm1987,Ilgisonis1999,Hameiri2004,Torasso2005,Hirota2006}, and a few of them employing the complete two-fluid model \cite{Elsasser1997,Spiess1999}. \\ \indent
A very useful apparatus for conducting stability analysis is the Hamiltonian description of  ideal fluid and plasma models.  The Hamiltonian framework, when adopting either a canonical description within the Lagrangian viewpoint or a noncanonical description within the Eulerian one, is a convenient framework for studying linearized dynamics and constructing functionals that can be exploited to establish stability criteria. Fluid and plasma criteria, such as the MHD energy principle and the Rayleigh criterion for shear flow,  ultimately exist because of the Hamiltonian form that can serve as a guide. \\ \indent
In this paper, we conduct formal stability analyses within the framework of a quasineutral two-fluid model with electron inertia, the so-called extended MHD (XMHD) model (e.g.\ see \cite{Lust1959,Kimura2014}). %, evidently, for the first time. 
 Attention has  been drawn to XMHD because of the recent discovery of its Hamiltonian structure  \cite{Abdelhamid2015} and its remarkable similarities with the Hamiltonian structure of HMHD \cite{Lingam2015,Lingam2016,Avignon2016}. We exploit this noncanonical Hamiltonian description of the model to employ the energy-Casimir (EC) and dynamically accessible (DA) methods \cite{Morrison1989,Morrison1990,Morrison1998} for deriving sufficient stability criteria upon constructing appropriate Lyapunov functionals. Moreover, using the action formalism developed in \cite{Charidakos2014} and \cite{Avignon2016} we examine the Lagrangian stability of the quasineutral two-fluid model by deriving the Hamiltonian of the corresponding linearized system in terms of Lagrangian displacements. Neglecting electron inertia, we derive a Hall MHD Lagrangian stability criterion that takes also into account the electron pressure contribution. Each one of the above stability methods has certain advantages and disadvantages which are discussed in detail in their respective sections. We can briefly say though that when applied under the same conditions, an ordering between them emerges from the dynamical point of view \cite{Andreussi2013}. The EC variations, being dynamically unconstrained, are more generic than the Lagrangian ones, which are generated through certain relations from arbitrary displacement vectors. In turn, the latter are more generic than the DA set of variations that are restricted by Hamiltonian dynamics. \\ \indent
%Although there are stability studies that take into account electron inertial effects e.g.\  \cite{Tassi2008}, to our knowledge there are no other similar results for three-dimensional and axisymmetric plasmas, utilizing the variety of methods included in the present work. 
The aim of this study is to provide a framework for formal stability analyses within a two-fluid description, which is more accurate and generic than that for MHD, staying though conceptually and formalistically as close as possible to MHD. In addition,  this work emphasizes that the Hamiltonian approach provides a unifying framework for studying equilibrium and stability employing the same principles.\\ \indent
The main ingredients of the Hamiltonian formulation of XMHD are, the Hamiltonian functional  \cite{Kimura2014,Abdelhamid2015}
%%%%%%
\begin{eqnarray}
\Hc&=&\frac{1}{2}\int_Vd^3x\,\left[\rho v^2+2\rho U(\rho)+B^2+d_e^2\frac{|\nb\times\bB|^2}{\rho}\right]\,,
\nn\\
&=&\frac{1}{2}\int_V d^3x\,\left[\rho v^2+2\rho U(\rho)+\bB\cdot\bB^*\right]\,,
 \label{hamiltonian}
\end{eqnarray} 
%%%%%%
where $V\subseteq \R^3$,  and the  noncanonical Poisson bracket \cite{Abdelhamid2015}, 
\begin{eqnarray}
\left\{F,G\right\}&=&\int_Vd^3x\,\big\{G_\rho\nb\cdot F_\bv - F_\rho\nb\cdot G_\bv
\label{poisson}\\
&+&\rho^{-1}\left(\nb\times\bv\right)\cdot\left(F_\bv \times G_\bv\right) \nn \\
&+& \rho^{-1}\bB^*\cdot\left[F_{\bv}\times\left( \nb\times G_{\bB^*}\right)-G_{\bv}\times \left(\nb\times F_{\bB^*}\right)\right] \nn \\
&-& d_i \rho^{-1}\bB^*\cdot\left[\left(\nb\times F_{\bB^*}\right)\times \left(\nb\times G_{\bB^*}\right)\right]\nn \\
&+& d_e^2\rho^{-1}\left(\nb\times \bv\right)\cdot \left[\left(\nb\times F_{\bB^*}\right)\times \left(\nb\times G_{\bB^*}\right)\right]\big\}\,, \nn
\end{eqnarray}
where $F_u:={\delta F}/{\delta u}$ denotes the functional derivative of $F$ with respect to the dynamical variable $u$.  The Poisson bracket of \eqref{poisson} is a generalization of that first given for MHD in \cite{morr-gre}.  Here the set of dynamical variables, say $\mathbf{u}$,  are the mass density $\rho$ the fluid velocity $\bv$ and the generalized magnetic field $\bB^*$ suggested in \cite{ling_morr_tass}, given by 
\begin{eqnarray}
\bB^*&=&\bB+d_e^2\nb\times\left(\frac{\nb\times\bB}{\rho}\right)\,. \label{B*}
\end{eqnarray}
The parameters $d_i$ and $d_e$ are the normalized ion and electron skin depths, respectively.
The  equations of motion for XMHD arising  from $\partial_t\mathbf{u}=\{\mathbf{u},\Hc\}$ are the following: 
\begin{eqnarray}
\partial_{t}\rho =&&-\nb\cdot\left(\rho \bv\right)\,,
\label{ConDen} \\
\partial_t \bv =&&\bv\times\bom-\nb\left(h+\frac{v^2}{2}+d_e^2\frac{|\bJ|^2}{2\rho^2}\right)+\frac{\bJ\times\bB^*}{\rho}\,, \label{mom_eq}\\
\partial_t\bB^* =&&\nb\times\left(\bv\times\bB^*-d_i\frac{\bJ\times \bB^*}{\rho} +d_e^2\frac{\bJ\times\bom}{\rho}\right) \label{ind_eq}\,,
\label{starB}
\end{eqnarray}
where $\bom:=\nb\times\bv$ and  $\bJ=\nb\times\bB$. \\ \indent
The degeneracy and explicit dependence of the noncanonical Poisson bracket on the dynamical variables $\mathbf{u}=(\rho,\bv,\bB^*)$, result in the emergence of topological constants of motion, called Casimirs, satisfying $\{F,\Cc\}=0$, $\forall F$. The presence of these invariants and their topological consequences, give rise to the EC and DA method. Exploiting these methodologies, we construct Lyapunov functionals suitable for establishing sufficient stability criteria without any reference to the dynamical equations: the perturbative procedure is implemented exclusively on the Hamiltonian level.\\ \indent
This paper is organized as follows: in Sec.~\ref{sec_II} we employ the EC method for studying the stability of axisymmetric XMHD equilibria. In this framework, several sufficient stability criteria are derived, concerning either special equilibria or special perturbations. In Sec.~\ref{sec_III} we find the dynamically accessible variations for the XMHD model, i.e.,  variations that keep the phase space trajectory on Casimir leaves. In addition, the second order, dynamically accessible variation of the Hamiltonian is utilized in order to establish a stability criterion for generic equilibria. Finally, in Sec.~\ref{sec_IV}, we compute the second order variation of the Lagrangian in a mixed Eulerian-Lagrangian framework and furthermore employ a Lagrange-Euler map to express the Lagrangian completely in terms of Eulerian coordinates. These results are used to construct the Hamiltonians for the linearized dynamics of the quasi-neutral two-fluid model and Hall MHD.

%%%%%%%%%%%%%%%%%%%%%%%%%%%%%
%%%%%%%%%%%%%%%%%%%%%%%%%%%%%
%%%%%%%%%%%%%%%%%%%%%%%%%%%%%

\section{Energy-Casimir stability of axisymmetric equilibria}
\label{sec_II}
In \cite{Kaltsas2018a}, we derived the equilibrium equations for helically symmetric and axisymmetric barotropic plasmas described by XMHD, using the EC principle. That principle can be extended to the computation of the second order variation which when evaluated on the EC equilibrium, denoted here as $\mathbf{u}_e$ is conserved by the linearized dynamics (e.g. \cite{Holm1985,Morrison1998}), and therefore a sufficient linear stability condition can be established by requiring that $\delta^2(\Hc-\sum_i\Cc_i)[\mathbf{u}_e,\delta\mathbf{u}]$ has definite sign. %Since $\delta^2(\Hc-\sum_i\Cc_i)[\mathbf{u}_e,\delta\mathbf{u}]$ provides a conserved norm for measuring linear deviations from equilibrium, we understand that EC stability implies linear and spectral stability since, the latter concerns just a special kind of perturbations which are included in the former. 
In general, however, the applicability of the EC method is not guaranteed since it requires a sufficient number of Casimir invariants in order to be established. This is the reason why in three-dimensional systems EC stability is usually not possible, other than special cases when  there exist some kind of Ertel's invariants, emerging usually due to entropy advection and  providing additional Casimirs \cite{Holm1985}. This would be the case also for XMHD if a baroclinic thermodynamic closure had been used. Ultimately the lack of Casimirs  was shown to be caused by the kind of degeneracy of the Poisson bracket in \cite{Morrison1998}. However, if a continuous spatial symmetry is present, the usual helicities are converted to infinite families of invariants in view of the symmetric decomposition of the fields, thus rendering the EC method applicable, as for example in \cite{Almaguer1988,amp1,Moawad2013,Andreussi2013,Andreussi2016} for the MHD model. One has to keep in mind though that this symmetric decomposition of the fields restricts the variations so as  to respect the geometrical symmetry of the system as well.

%%%%%%%%%%%%%%%%%%%%%%%%%%%%%
%%%%%%%%%%%%%%%%%%%%%%%%%%%%%

\subsection{Axisymmetric XMHD energy-Casimir functional}

The axisymmetric velocity and magnetic fields can be Helmholtz-decomposed as follows
%%%%%%%%
\begin{eqnarray}
\bv=&&r v_\phi \nb\phi+\nb\chi\times\nb\phi+\nb\Upsilon \,, \\
\bB=&&r B_\phi \nb\phi+\nb\psi\times\nb\phi \,,
\end{eqnarray} 
%%%%%%%%%
inducing a similar form for the generalized magnetic field $\bB^*$.
From Eqs.~(4.10)--(4.13) in \cite{Kaltsas2018a}  we can easily obtain the following axisymmetric Casimirs
%%%%%%%%
\begin{eqnarray}
\Cc_1&=&\int_Dd^2x\,(r^{-1}B^*_\phi+\gamma \Omega )\Fc(\psi^*+\gamma r v_\phi)\,,\\
\Cc_2&=&\int_Dd^2x\,(r^{-1}B^*_\phi+\mu \Omega )\Gc(\psi^*+\mu r v_\phi)   \,, \\
\Cc_3&=&\int_Dd^2x\, \rho  \Mc(\psi^*+\gamma r v_\phi) \,, \\
\Cc_4&=&\int_Dd^2x\,  \rho \Nc(\psi^*+\mu r v_\phi)  \,, 
\end{eqnarray}
%%%%%%%%
where $\Omega:=(\nb\times\bv_\perp)\cdot\nb\phi$ with $\bv_\perp:=\nb\chi\times\nb\phi+\nb\Upsilon$ and $\psi^*=\psi-d_e^2\rho^{-1}\Delta^*\psi$, $B_\phi^*=B_\phi-d_e^2 r \nb \cdot \left[r^{-2}\rho^{-1}\nb(rB_\phi)\right]$, with $\Delta^*:=r^2\nb\cdot(r^{-2}\nb)$ being the so-called Shafranov operator. The parameters $\gamma$ and $\mu$ are $(\gamma,\mu)=(\gamma_+,\gamma_-)$ where $\gamma_\pm = (d_i \pm \sqrt{d_i^2 + 4d_e^2} )/2$. The axisymmetric Hamiltonian is given by
%%%%%%%%%%
\begin{eqnarray}
\Hc=\int_Dd^2x\, \bigg( \rho\frac{v_\phi^2}{2}+\rho \frac{|\nb\chi|^2}{2r^2}+\rho\frac{|\nb\Upsilon|^2}{2}\nn \\
+\rho[\Upsilon,\chi]+\rho U(\rho)+\frac{B_\phi^* B_\phi}{2}+\frac{\nb\psi^*\cdot\nb\psi}{2r^2}\bigg)\,. \label{axis_ham}
\end{eqnarray}
The vanishing of the first order variation of the EC functional, i.e.,  $\delta\Hc_C=\delta(\Hc-\sum_i \Cc_i)=0$, yields the EC equilibrium equations, given by Eqs.~(4.25)--(4.31) of \cite{Kaltsas2018a} with $\ell=0,n=-1$ therein, which can be written in a Grad-Shafranov-Bernoulli form (see Eqs.~(5.1)--(5.4) in the same reference).  In this case, $\delta\Hc_C$ assumes the form 
%%%%%%%%%%%%
\begin{widetext}
\begin{eqnarray}
&&\delta\Hc_C=\int_Dd^2x\,\Bigg\{ \bigg[h(\rho)-\Mc-\Nc+\frac{v_\phi^2}{2}+\frac{|\bv_\perp|^2}{2}+\frac{d_e^2}{2r^2\rho^2}\left((\Delta^*\psi)^2+|\nb(rB_\phi)|^2\right)\bigg]\delta\rho+
\left[B_\phi-r^{-1}(\Fc+\Gc)\right]\delta B_\phi^*\nn \\
&&+\left[\rho \bv_\perp-\gamma\nb\Fc\times\nb\phi-\mu\nb\Gc\times\nb\phi\right]\cdot\delta \bv_\perp +\left[\rho v_\phi-\gamma r(r^{-1}B_\phi^*+\gamma \Omega)\Fc'-\mu r(r^{-1}B_\phi^*+\mu \Omega) \Gc'-\gamma r\rho \Mc'-\mu r \rho \Nc'\right]\delta v_\phi\nn\\
&&-\left[r^{-2}\Delta^*\psi+(r^{-1}B_\phi^*+\gamma \Omega)\Fc'+(r^{-1}B_\phi^*+\mu \Omega)\Gc'+\rho\Mc'+\rho\Nc'\right]\delta \psi^*\Bigg\}\,. \label{deltaF}
\end{eqnarray}
\end{widetext}

%%%%%%%%%%%%%%%%%%%%%%%%%%%%%
%%%%%%%%%%%%%%%%%%%%%%%%%%%%%

\subsection{Second order variation}

The expressions into the square brackets in \eqref{deltaF} vanish on the EC equilibrium solution; therefore the second order variation would involve only first order variations of the fields. After some manipulations $\delta^2\Hc_C[\mathbf{u}_e,\delta \mathbf{u}]$ can be written in the following form:
%%%%%%%
%\begin{widetext}
\begin{eqnarray}
&&\delta^2\Hc_C[\mathbf{u}_e;\delta \mathbf{u}]=\int_Dd^2x\, \bigg\{\frac{d_e^2}{\rho r^2}|\nb(r\delta B_\phi)|^2+ \frac{|\nb\delta\psi|^2}{r^2}\nn \\ 
&&\quad +\frac{d_e^2r^2}{\rho}\left[\nb\cdot\left(r^{-2}\nb\delta\psi\right)\right]^2+\rho\left(\delta v_\phi+\rho^{-1}v_\phi\delta\rho\right)^2\nn\\
&&\quad  +\rho\big|\delta \bv_\perp +\rho^{-1}\bv_\perp\delta\rho\big|^2-2\frac{d_e^2}{r^2\rho}\nb(\delta\Fc+\delta\Gc)\cdot\nb(r\delta B_\phi)\nn\\
&&\quad +2 \frac{d_e^2}{r^2\rho^2}\nb(\delta \Fc+\delta\Gc)\cdot\nb(rB_\phi)\delta\rho\nn\\
&&\quad -2[(\gamma\nb\delta\Fc+\mu\nb\delta\Gc)\times\nb\phi]\cdot\delta\bv_\perp\bigg\}+\Qc\,,\label{d2Hc}
%\Mc'\left[\delta\rho-(\delta\psi^*+\gamma r\delta v_\phi)\right]^2+\Nc'\left[\delta\rho-(\delta\psi^*+\mu r\delta v_\phi)\right]^2\nn \\
\end{eqnarray}
%\end{widetext}
where 
%$\delta\varphi:=\psi^*+\gamma r v_\phi$, $\delta\xi:=\psi^*+\mu r v_\phi$, 
%$\delta\Omega=(\nb\times\delta\bv_\perp )\cdot\nb\phi$
%and 
\begin{equation}
\Qc=\int_Dd^2x\, (\delta B_\phi\; \delta\varphi\;\delta\xi\;\delta\rho)\,\Ac\,(\delta B_\phi\;\delta\varphi\;\delta\xi\;\delta\rho)^T\,, \label{Q1}
\end{equation}
with
\begin{eqnarray}
\Ac=
\begin{pmatrix}
1&& A_{\varphi B_\phi} && A_{\xi B_\phi}&& 0\\ 
A_{\varphi B_\phi}&& A_{\varphi \varphi} && 0 &&A_{\varphi \rho} \\
A_{\xi B_\phi}&& 0 && A_{\xi \xi} && A_{\xi \rho}\\
0&& A_{\varphi \rho}&& A_{\xi \rho}  &&A_{\rho \rho}
\end{pmatrix}\, , \label{stamatrix}
\end{eqnarray}
and the  elements of $\cal{A}$  given explicitly by 
\begin{eqnarray}
%A_{11}=&&1\,,\\
A_{\varphi \varphi}=&&-\left(r^{-1}B_\phi^*+\gamma\Omega\right)\Fc''-\rho\Mc''\\
A_{\xi \xi}=&&-\left(r^{-1}B_\phi^*+\mu\Omega\right)\Gc''-\rho\Nc'' \\
A_{\varphi B_\phi }=&& -r^{-1}\Fc'\,,\quad A_{\xi B_\phi}=-r^{-1}\Gc'\,,\\
A_{\varphi \rho}=&&-\Mc'\,,\quad A_{\xi \rho}=-\Nc'\\
A_{\rho \rho}=&&\rho^{-1}\bigg[c_s^2-v_\phi^2-|\bv_\perp|^2\\
&&\hspace{-0.5cm}-\frac{d_e^2}{\rho^2}\left(r^2\left[\nb\cdot\left(r^{-2}\nb\psi\right)\right]^2+r^{-2}\big|\nb(rB_\phi)\big|^2\right)\bigg]\,,
\nonumber
\end{eqnarray}
where $c_s^2:=\rho h'(\rho)$. In deriving  \eqref{d2Hc}, we  integrated by parts, omitted  the surface integrals,  and completed squares in terms involving the mass density and velocity field variations.\\ \indent
For $\Qc$ alone to be positive definite, the matrix $\Ac$ has to be positive definite,  which is equivalent to the requirement that the principal minors of $\Ac$ satisfy
%%%%%%
\begin{eqnarray}
&&A_{\varphi \varphi}-A_{\varphi B_\phi}^2 > 0\,, \label{Apd_1}\\
&& A_{\xi \xi}(A_{\varphi \varphi}-A_{\varphi B_\phi}^2)-A_{\varphi \varphi}A_{\xi B_\phi}^2> 0\,, \label{Apd_2}\\
&&A_{\rho \rho}\left[A_{\xi \xi}(A_{\varphi \varphi}-A_{\varphi B_\phi}^2)-A_{\varphi \varphi}A_{\xi B_\phi}^2\right]\nn\\
&&\qquad +(A_{\varphi B_\phi}A_{\xi \rho}-A_{\xi B_\phi}A_{\varphi \rho})^2\nn\\
&&\qquad \qquad -A_{\xi \xi}A_{\varphi \rho}^2-A_{\varphi \varphi}A_{\xi \rho}^2> 0\,. \label{Apd_3}
\end{eqnarray}
%%%%%%%%%%%%
%For the conditions \eqref{Apd_1}--\eqref{Apd_2} to hold, it is necessary that $A_{\varphi \varphi}> 0$ and  $A_{\xi \xi}> 0$.%,  which could be related to the suppression of current driven instabilities. This can be understood upon noticing that one of the ``Euler-Lagrange'' equations resulting from the variational principle \eqref{deltaF} relates the toroidal current density with the quantities $A_{\varphi \varphi}$ and $A_{\xi \xi}$.
However, $\Qc>0$ does not imply stability because there are several indefinite terms in $\delta^2\Hc_C$. More precisely, the first five terms in $\delta^2\Hc_C$ are always non-negative, with the magnetic terms expressing the magnetic field line bending, while the other two terms contain kinetic energy and compressional contributions of the perturbation. These kinetic-compressional terms constitute an example of the non-separability of energies mentioned in the introduction, rendering the resulting stability conditions sufficient but not necessary. The nonseparability is even more severe, since kinetic and potential energy contributions are intertwined also via other terms in $\delta^2\Hc_C$ reflecting the fact that in the two-fluid framework, the coupling between flows and magnetic fields is more complicated.  In particular, what really makes life difficult, are the last three terms into the curly bracket in \eqref{d2Hc} because they are clearly sources of indefiniteness, a characteristic that has been identified  in previous EC stability analyses of similar models  \cite{Tassi2008,Tassi2012}, and can potentially be related to linear instability or the presence of Negative Energy Modes (NEMs). Both can lead to disastrous destabilization and loss of confinement. In order to remove the indefiniteness, we can  eliminate or conflate these ``problematic'' terms into other terms in view of certain constraints imposed on the variations $\delta B_\phi^*$ and $\delta\Omega$ or by considering special equilibria. %We delineate various possibilities for removing the indefiniteness of $\delta^2\Hc_C$ in the subsections below.

%%%%%%%%%%%%%%%%%%%%%%%%%%%%%
%%%%%%%%%%%%%%%%%%%%%%%%%%%%%

\subsection{Special equilibria}

%%%%%%%%%%%%%%%%%%%%%%%%%%%%%

\subsubsection{Extended MHD}

For purely toroidal flow and current, i.e.\  $\Fc'=\Gc'=0$, it is clear that $\Qc > 0$ implies  $\delta^2\Hc_C>0$.   For our special class of equilibria, we have $A_{\varphi B_\phi}=A_{\xi B_\phi}=0$, and consequently conditions \eqref{Apd_1}--\eqref{Apd_3} yield
%%%%%%%%%5
\begin{eqnarray}
&&\Mc''< 0\,, \quad \Nc''< 0\,, \label{stacond_xmhd_tor_rot12}\\
&&\Mc''\Nc''\bigg[c_s^2-v_\phi^2-\frac{d_e^2}{\rho^2}\left(r^2\left[\nb\cdot\left(r^{-2}\nb\psi\right)\right]^2\right)\bigg]\nn\\
&&\qquad +\Mc''(\Nc')^2+\Nc''(\Mc')^2 > 0\,. \label{stacond_xmhd_tor_rot3}
\end{eqnarray}
The first two conditions imply that $\Mc$ and $\Nc$ must  be concave functions. For the condition \eqref{stacond_xmhd_tor_rot3} to be satisfied, the quantity inside the square bracket must necessarily be  positive, that is, the toroidal velocity modified by an electron inertial correction has to be lower than the speed of sound, thus preventing shock formation.% Note also that \eqref{stacond_xmhd_tor_rot12} comes from $A_{\varphi \varphi}\geq 0$ and  $A_{\xi \xi}\geq 0$ which could be related to the suppression of current driven instabilities. This can be understood upon noticing that one of the ``Euler-Lagrange'' equations resulting from the variational principle \eqref{deltaF} relates the toroidal current density with the quantities $A_{\varphi \varphi}$ and $A_{\xi \xi}$.

%%%%%%%%%%%%%%%%%%%%%%%%%%%%%

\subsubsection{Hall MHD}

In the limit $d_e\rightarrow 0$, $\mu \rightarrow 0$ as well,  and  there is only one indefinite term in \eqref{d2Hc}, which can be removed upon selecting $\Fc'=0$. In this case, the flow is purely toroidal, but there is poloidal current created by the electrons. From \eqref{Apd_1}--\eqref{Apd_3}, we obtain the following sufficient stability conditions:
\begin{eqnarray}
&&\Mc''< 0\,, \label{stacond_hall_tor_rot1}\\
&& r^{-2}\Gc\Gc''+\rho \Nc''+r^{-2}(\Gc')^2 < 0\,, \label{stacond_hall_tor_rot2}\\
&&\left[\Mc''(c_s^2-v_\phi^2)+(\Mc')^2\right]\left[r^{-2}\Gc\Gc''+\rho\Nc''+r^{-2}(\Gc')^2\right]\nn\\
&&\hspace{2.2cm}+\rho\Mc''(\Nc')^2 > 0\,. \label{stacond_hall_tor_rot3}
\end{eqnarray}
The conditions above necessarily entail $c_s^2-v_\phi^2> 0$. This special case is interesting because the stability condition is expressed explicitly in terms of equilibrium quantities, and  furthermore,  it allows us to  study  the stability of nontrivial equilibria. For this reason, we proceed by constructing a Hall MHD equilibrium with purely toroidal rotation and applying  the criterion \eqref{stacond_hall_tor_rot1}--\eqref{stacond_hall_tor_rot3}. From  $\delta \Hc_c=0$ (see \eqref{deltaF}), setting $d_e=0$ and imposing $\bv_\perp=\delta \bv_\perp=0$ we can easily extract the equilibrium equations of interest. These are
%%%%%%%%%
\begin{eqnarray}
&&\Delta^*\psi+\Gc\Gc'(\psi)+\rho \frac{\varphi-\psi}{d_i^2}+r^2\rho\Nc'(\psi)=0\,, \label{hall_tor_rot_1}\\
&&h(\rho)=\Mc(\varphi)+\Nc(\psi)-\frac{v_\phi^2}{2}\,,\label{hall_tor_rot_2}\\
&&B_\phi=r^{-1}\Gc(\psi)\,,\; v_\phi=d_i r\Mc'(\varphi)\,,\label{hall_tor_rot_3}\\
&&\varphi-d_i^2r^2\Mc'(\varphi)=\psi\,, \label{hall_tor_rot_4}
\end{eqnarray}
%%%%%%%%%%%%
where we have used the definition of $\varphi$ to write $v_\phi=\frac{\varphi-\psi}{d_i r}$. Additionally,  we consider the following nonlinear ansatz for the free functions $\Gc$, $\Mc$ and $\Nc$:
\begin{eqnarray}
\Gc=&&g_0+g_1\psi+\frac{1}{2}g_2\psi^2+\frac{1}{3}g_3\psi^3\,,\nn\\
\Mc=&&m_0+m_1\varphi+\frac{1}{2}m_2\varphi^2+\frac{1}{3}m_3\varphi^3\,,\nn\\
\Nc=&&n_0+n_1\psi+\frac{1}{2}n_2\psi^2+\frac{1}{3}n_3\psi^3\,, \label{ansatz}
\end{eqnarray}
%%%%%%%%%%
%and we choose the negative solution branch of Eq. \eqref{hall_tor_rot_4}. 
We set $m_1=0$,  which implies that there exists a solution to \eqref{hall_tor_rot_4} for which  $\varphi=0$ wherever $\psi=0$; therefore the two flux functions satisfy the same boundary condition. We consider an adiabatic equation of state, i.e.,  $h(\rho)=\Gamma/(\Gamma-1)p_1\rho^{\Gamma-1}$, where $\Gamma=5/3$ is the adiabatic index and $p_1$ is a constant. Then, Eq.~\eqref{hall_tor_rot_1} was solved numerically, using  finite differences and a simple SOR iterative solver, on an up-down poloidally asymmetric domain with a prescribed diverted boundary having a lower x-point and tokamak pertinent values for the free parameters. \\ \indent
%%%%%%%%%%%%%%
\begin{figure}[h]
\includegraphics[ scale=0.35]{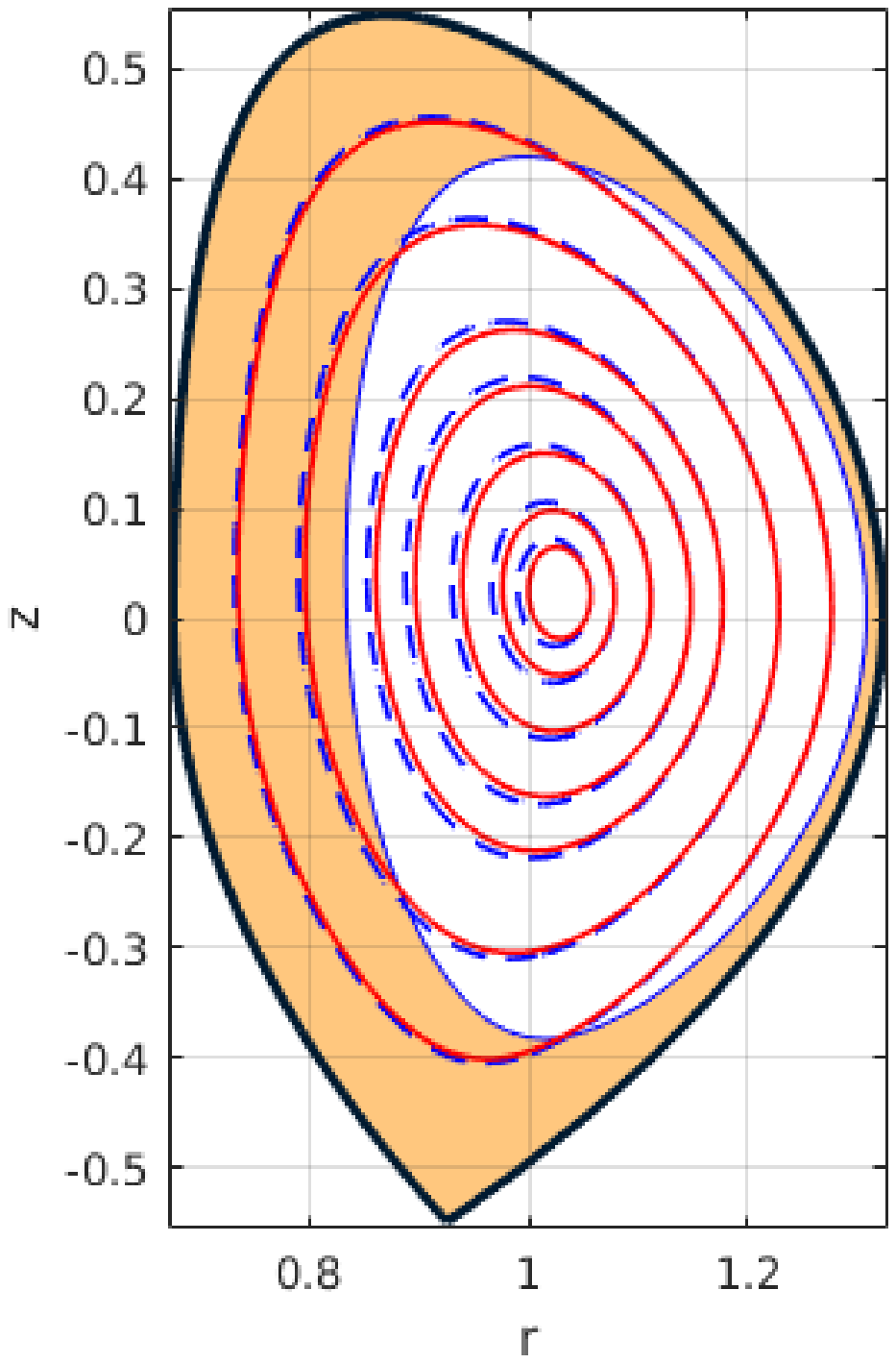} \includegraphics[ scale=0.35]{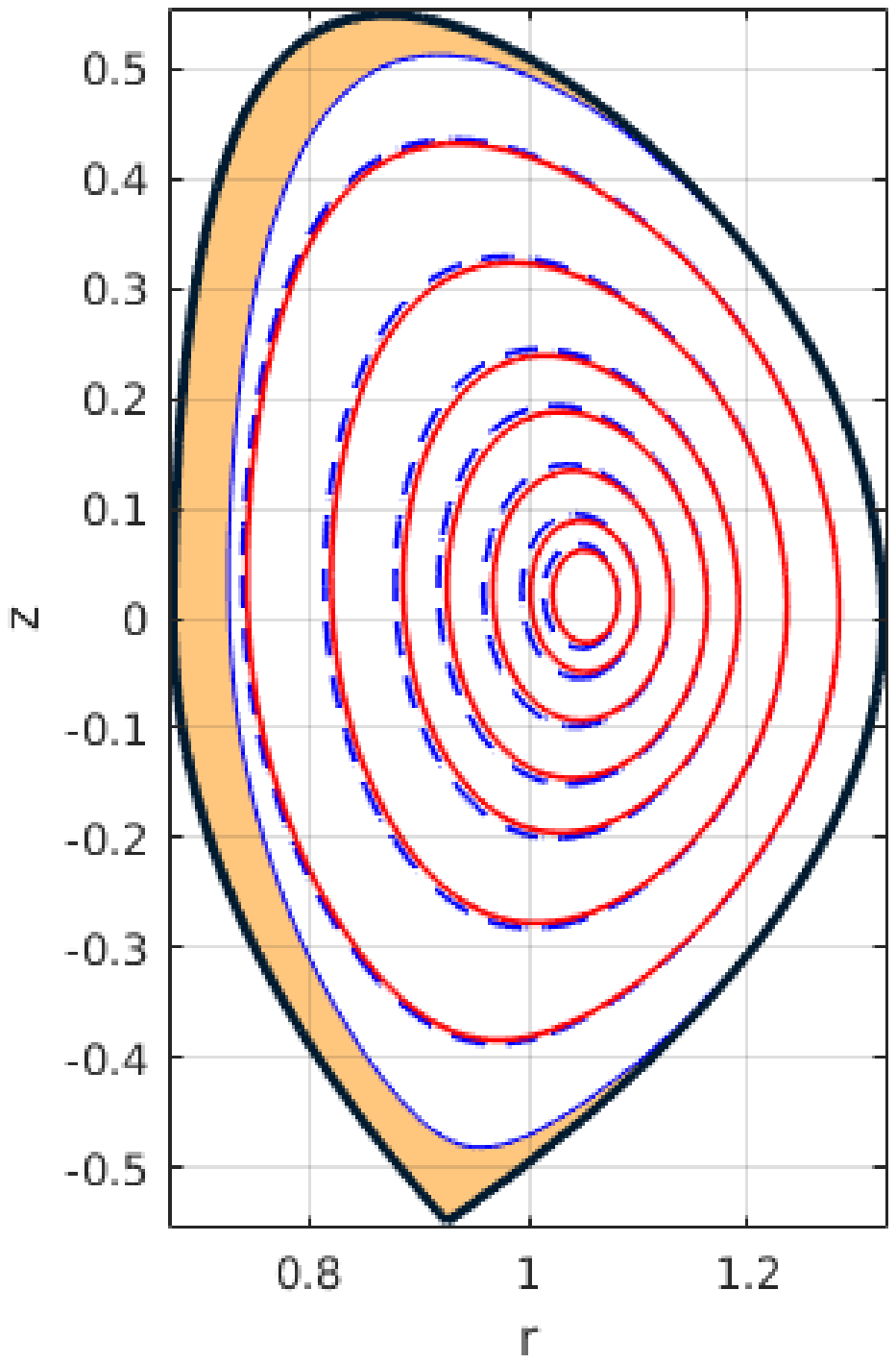}\\
\caption{The stability diagrams for two ITER-like equilibria with maximum $\beta\sim 2\%$ (left) and $\sim 20\%$ (right). In the coloured regions all three conditions \eqref{stacond_hall_tor_rot1}--\eqref{stacond_hall_tor_rot3} are satisfied. The Hall parameter is $d_i=0.04$ in both cases. Solid red lines represent the magnetic surfaces, while the dashed blue ones are surfaces of constant angular velocity.} \label{fig_0}
\includegraphics[ scale=0.35]{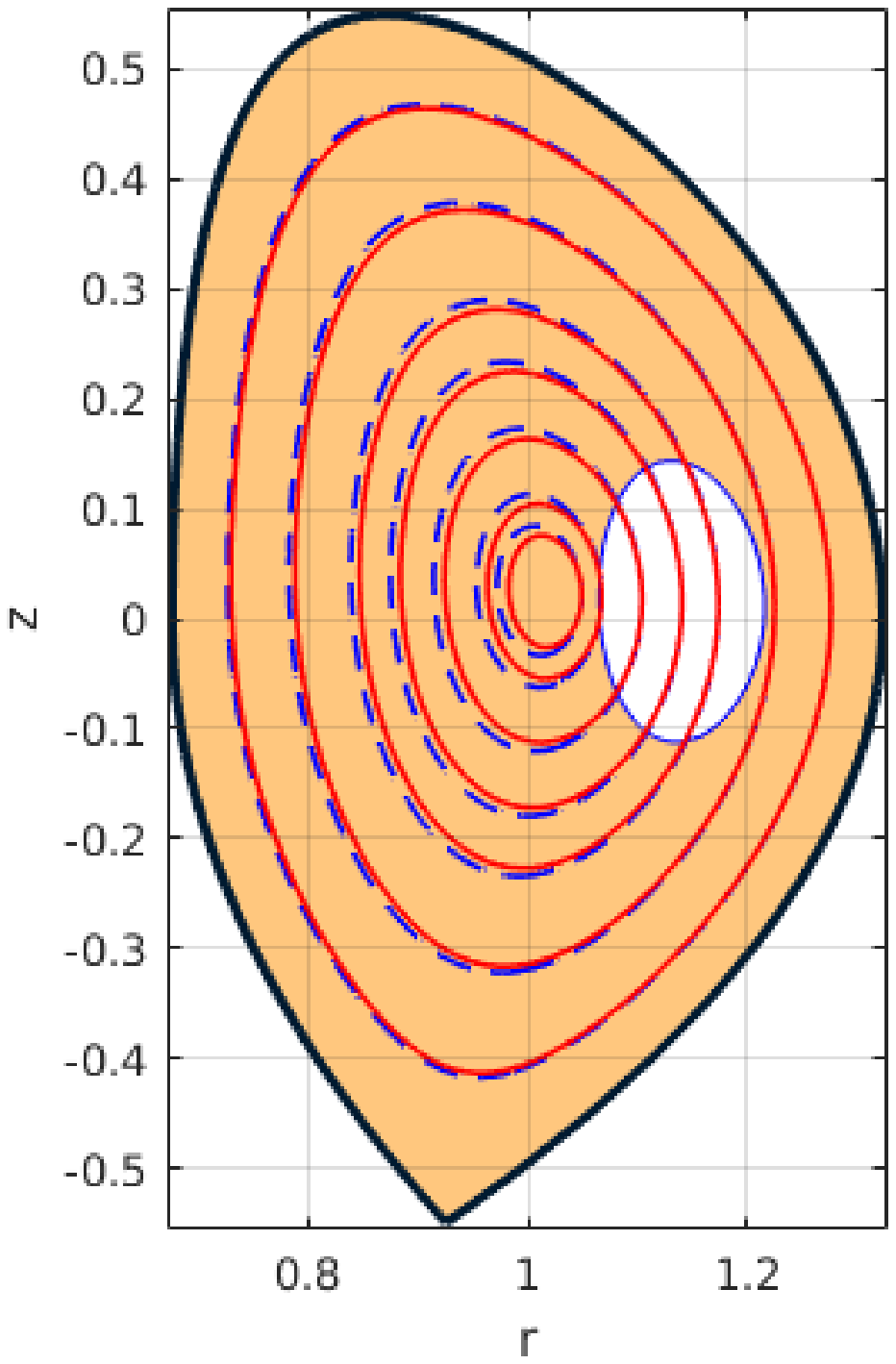} \includegraphics[scale=0.35]{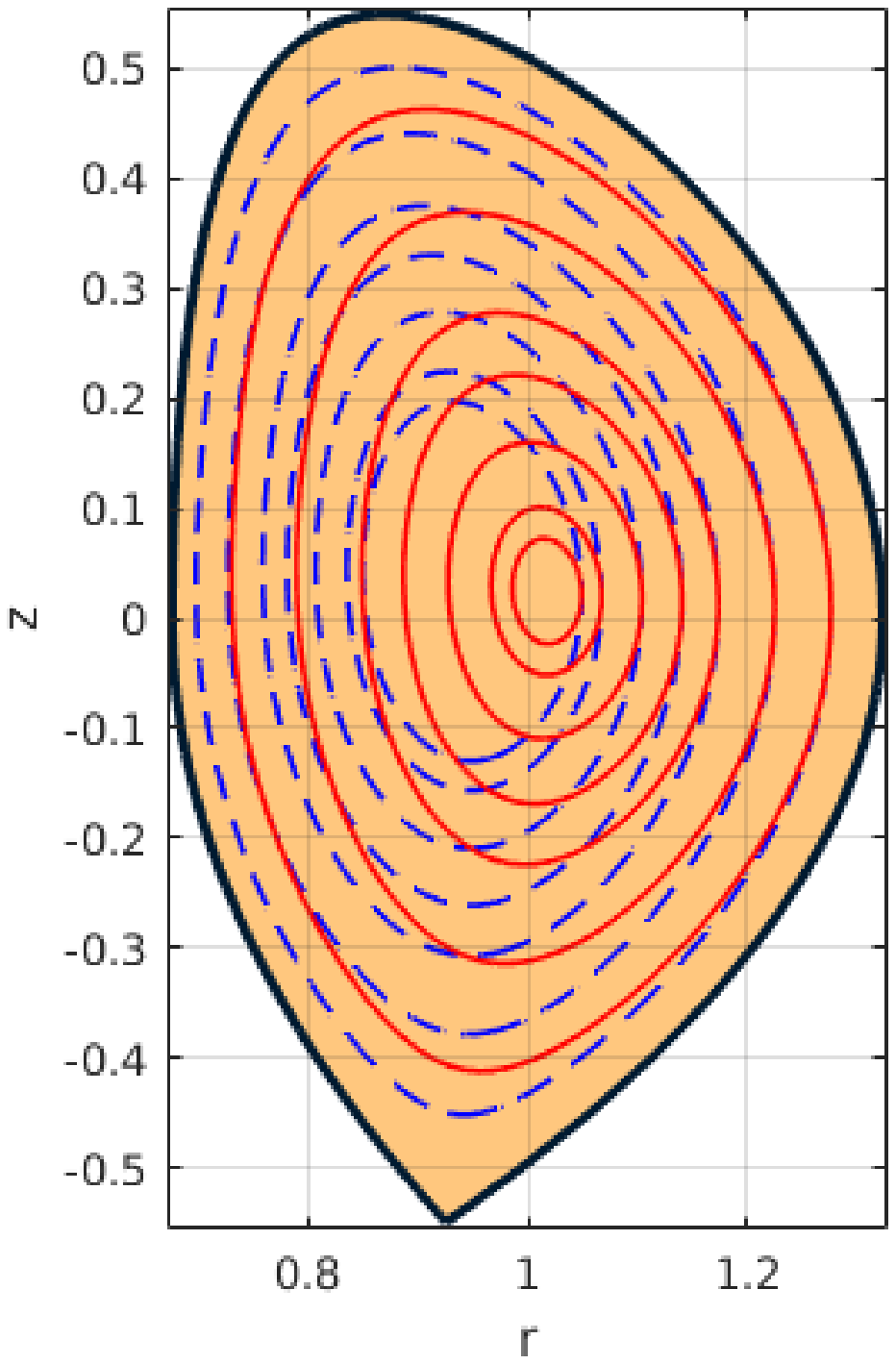}
\caption{The stability diagrams for equilibria with maximum $\beta\sim 0.8\%$ with $d_i=0.04$ (left) and $d_i=0.24$ (right). While for $d_i=0.04$ there is a hole within which \eqref{stacond_hall_tor_rot3} is not satisfied, increasing $d_i$ results in a completely stable configuration, under EC variations. }\label{fig_1}
\end{figure}
%%%%%%%%%%%%
It is not difficult to adjust the free parameters in \eqref{ansatz} to make conditions \eqref{stacond_hall_tor_rot1}--\eqref{stacond_hall_tor_rot2} to be satisfied everywhere in the plasma.  However, when it comes to \eqref{stacond_hall_tor_rot3} we observe that for   $\beta> 1\%$,  the condition is satisfied only within a narrow annular region, wider on the high field side and narrower on the low field side. For $\beta > 10\%$, this region is even narrower forming a thin layer spreading across the high field side only (Fig. \ref{fig_0}). For \  $\beta<1 \%$,  we were able to find equilibria that satisfy all three conditions \eqref{stacond_hall_tor_rot1}--\eqref{stacond_hall_tor_rot3} all over the computational domain. This indicates that condition \eqref{stacond_hall_tor_rot3} is potentially related with the stabilization of pressure driven modes. To capture the influence of the Hall parameter $d_i$ on stability, we considered an equilibrium with $d_i=0.04$ where all three stability conditions are satisfied everywhere outside a small region near the core. Then, we increased gradually $d_i$, observing that this region was continuously shrinking until it disappeared for $d_i=0.24$. Thereby, we conclude that upon increasing $d_i$, the stability properties may be improved (see Fig.~\ref{fig_1}). We also corroborated that if we include the linear term in $\Mc$, which is related to rigid rotation and therefore being intrinsically destabilizing, shrinks the ``stable'' region towards the high field side.  In closing, we  underline that an equilibrium that fails to satisfy the stability conditions is not necessarily unstable,  because the criteria we derived are only sufficient.

%%%%%%%%%%%%%%%%%%%%%%%%%%%%%
%%%%%%%%%%%%%%%%%%%%%%%%%%%%%

\subsection{Conditional stability (constrained variations)}
As mentioned earlier, the indefiniteness in $\delta^2\Hc_C$ comes from the terms in \eqref{d2Hc} containing $\delta \Fc$ and $\delta \Gc$ and multiplied by $\nb\times\delta \bv_\perp$, $\delta B_\phi$ and $\delta \rho$. Hence, a simple way to get rid of the indefiniteness is to assume $\delta\rho=\delta B_\phi=\nb\times\delta\bv_\perp=0$. However, such a severe restriction of the permitted perturbations should be justified on physical grounds. Another possibility is to assume only $\delta\rho=0$, which can be justified by the fact that incompressible variations are considered to be the most dangerous ones and then try to eliminate the explicit appearance of $\delta B_\phi$ and $\delta\bv_\perp$ into $\delta^2\Hc_C$ by other means.  A way to do so is to partially minimize the functional \eqref{d2Hc} with respect to $\delta\bv_\perp$ and $\delta B_\phi$. This is a standard procedure to obtain simplified forms of the Lyapunov functional and improved stability criteria (e.g.\  see \cite{Newcomb1960,Hameiri1998,Hameiri2003,Andreussi2013}). The minimization can be realized upon considering $\delta^2\Hc_C$ as a functional of the variations $\delta \mathbf{u}$ and require its variation with respect to $\delta B_\phi$ and $\delta\bv_\perp$ to vanish. The resulting Euler-Lagrange equations 
%%%%%%%%%% 
\begin{eqnarray}
\delta B_\phi&=&r^{-1}(\delta\Fc+\delta \Gc)\,,\label{dBphi_minimizer}\\
\delta\bv_\perp&=&-\bv_\perp\delta\rho+(\gamma \nb\delta\Fc+\mu\nb\delta\Gc)\times\nb\phi\,,\label{dvp_minimizer}
\end{eqnarray}
are indeed minimizers of the functional, since the second variation with respect  to $\delta B_\phi$ and $\delta \bv_\perp$ is positive definite. Henceforth, we set $\delta\rho=0$.
%%%%%%%%%%
Upon substituting Eqs.~\eqref{dBphi_minimizer}--\eqref{dvp_minimizer} into \eqref{d2Hc} we find
%%%%%%%%%%%
%\begin{widetext}
\begin{eqnarray}
&&\delta^2\tilde{\Hc}_C=\int_Dd^2x\,\bigg\{\frac{|\nb\delta\psi|^2}{r^2}+\frac{d_e^2r^2}{\rho}\left[\nb\cdot\left(r^{-2}\nb\delta\psi\right)\right]^2 \nn
\\&&
+\rho(\delta v_\phi)^2+r^{-2}(\delta\Fc-\delta\Gc)^2\bigg\}+\tilde{\Qc}
\end{eqnarray}
%\end{widetext}
and therefore $\tilde{\Qc}> 0$ implies stability. We have
%%%%%%%%%%%%%
\begin{eqnarray}
\tilde{\Qc}&=&\int_Dd^2x\, \left[A_{\varphi\varphi}(\delta\varphi)^2+A_{\xi\xi}(\delta\xi)^2\right]\nn\\
&&-\int_Dd^2x\,\bigg\{\frac{\gamma^2+d_e^2}{r^2\rho}\big[(\Fc')^2|\nb\delta\varphi|^2+2\Fc'\delta\varphi\nb\Fc'\cdot\nb\delta\varphi\big]\nn\\
&&+\frac{\mu^2+d_e^2}{r^2\rho}\big[(\Gc')^2|\nb\delta\xi|^2
+2\Gc'\delta\xi\nb\Gc'\cdot\nb\delta\xi\big]\nn\\
&&+\left[2r^{-2}(\Fc')^2+\frac{\gamma^2+d_e^2}{\rho r^2}|\nb\Fc'|^2\right](\delta\varphi)^2\nn\\
&&+\left[2r^{-2}(\Gc')^2+\frac{\mu^2+d_e^2}{\rho r^2}|\nb\Gc'|^2\right](\delta\xi)^2\bigg\}\,. \label{Q3}
\end{eqnarray}
%%%%%%%%%%%
%%%%%%
Following \cite{Holm1985}, let us define the vectors $\mathbf{k}_\varphi:=\nb\delta\varphi/\delta\varphi$, $\mathbf{k}_\xi:=\nb\delta\xi/\delta\xi$. In view of this definition, we can write \eqref{Q3} in diagonal form with
%%%%%%%%%%%%
\begin{eqnarray}
\tilde{A}_{\varphi\varphi}=&&-\left(r^{-1}B_\phi^*+\gamma\Omega\right)\Fc''-\rho\Mc''-2r^{-2}(\Fc')^2\nn\\
&&\hspace{-1cm}-\frac{\gamma^2+d_e^2}{\rho r^2}\left[|\nb\Fc'|^2+ (\Fc')^2|\mathbf{k}_\varphi|^2+\mathbf{k}_\varphi\cdot\nb(\Fc')^2\right]\,,\\
\tilde{A}_{\xi\xi}=&&-\left(r^{-1}B_\phi^*+\mu\Omega\right)\Gc''-\rho\Nc'' -2r^{-2}(\Gc')^2\nn\\
&&\hspace{-1cm}-\frac{\mu^2+d_e^2}{\rho r^2}\left[|\nb\Gc'|^2+ (\Gc')^2|\mathbf{k}_\xi|^2+\mathbf{k}_\xi\cdot\nb(\Gc')^2\right]\,.
%\tilde{A}_{\varphi\rho}=&&-\Mc'+\frac{\gamma^2+d_e^2}{r^2\rho^2}\Fc'\left(\Fc''|\nb\varphi|^2+\mathbf{k}_\varphi\cdot\nb\Fc\right)\,,\\
%\tilde{A}_{\xi\rho}=&&-\Nc'+\frac{\mu^2+d_e^2}{r^2\rho^2}\Gc'\left(\Gc''|\nb\xi|^2+\mathbf{k}_\xi\cdot\nb\Gc\right)\,,\\
%\tilde{A}_{\rho\rho}=&&\rho^{-1}\bigg[c_s^2-v_\phi^2-|\bv_\perp|^2\nn\\
%&&\hspace{-0.5cm}-\frac{d_e^2}{\rho^2}\left(r^2\left[\nb\cdot\left(r^{-2}\nb\psi\right)\right]^2+r^{-2}\big|\nb(rB_\phi)\big|^2\right)\bigg]\,.
\end{eqnarray}
%%%%%%%%%%%%%%%%%
Invoking the Cauchy-Schwartz inequality, it is clear that the following conditions
%%%%%%%%%%%%
\begin{eqnarray}
&&-\left(r^{-1}B_\phi^*+\gamma\Omega\right)\Fc''-\rho\Mc''-2r^{-2}(\Fc')^2\nn\\
&&\qquad -\frac{\gamma^2+d_e^2}{\rho r^2}\left[|\nb\Fc'|^2+(\mathcal{F}' )^2 |\mathbf{k}_\varphi|^2+|\mathbf{k}_\varphi| |\nabla (\mathcal{F}')^2| \right]\nn \\
&&\qquad \equiv a_\varphi |\mathbf{k}_\varphi|^2+b_\varphi|\mathbf{k}_\varphi|+c_\varphi > 0\,, \label{App_C-S}\\
&&\hspace{1cm}-\left(r^{-1}B_\phi^*+\mu\Omega\right)\Gc''-\rho\Nc'' -2r^{-2}(\Gc')^2\nn\\
&&\hspace{1cm}-\frac{\mu^2+d_e^2}{\rho r^2}\left[|\nb\Gc'|^2+(\mathcal{G}' )^2 |\mathbf{k}_\xi|^2+|\mathbf{k}_\xi| |\nabla (\mathcal{G}')^2|\right]\nn \\
&&\qquad \equiv a_\xi |\mathbf{k}_\xi|^2+b_\xi|\mathbf{k}_\xi|+c_\xi > 0\,, \label{Axx_C-S}
\end{eqnarray}
%%%%%%%%%%%
are sufficient for $\tilde{A}_{\varphi\varphi}> 0$ and $\tilde{A}_{\xi\xi} > 0$ and therefore for $\tilde{\Qc}> 0$. The two polynomials in $|\mathbf{k}_\varphi|$ and $|\mathbf{k}_\xi|$ must have at least one real positive root. Given that $a_\varphi<0$, $b_\varphi<0$  and $a_\xi<0$, $b_\xi<0$, we understand that one root will be always negative; thus, in order for the second one to be positive, the products of the roots given by $c_\varphi/a_\varphi$, $c_\xi/a_\xi$, must be negative. Therefore,  we conclude that the conditions under which there exist exactly one real positive root for each polynomial are 
%%%%%%%%5
\begin{eqnarray}
c_\varphi&:=&-\left(r^{-1}B_\phi^*+\gamma\Omega\right)\Fc''-\rho\Mc''\nn\\
&&-2r^{-2}(\Fc')^2-\frac{\gamma^2+d_e^2}{\rho r^2}|\nb\Fc'|^2> 0\,, \label{cphi_condition}\\
c_\xi&:=&-\left(r^{-1}B_\phi^*+\mu\Omega\right)\Gc''-\rho\Nc'' \nn\\
&&-2r^{-2}(\Gc')^2-\frac{\mu^2+d_e^2}{\rho r^2}|\nb\Gc'|^2 > 0\,. \label{cxi_condition}
\end{eqnarray}
Now in view of \eqref{cphi_condition}--\eqref{cxi_condition} the two polynomials are also positive in the domain $0\leq |\mathbf{k}_\varphi|< k_\varphi^{+}$, $0\leq|\mathbf{k}_\xi|< k_\xi^{+}$, where $k_\varphi^{+}$ and $k_\xi^{+}$ are the real positive roots of the polynomials in \eqref{App_C-S} and \eqref{Axx_C-S}, respectively. This is true, since they do not change sign within this domain and furthermore they are positive for $|\mathbf{k}_\varphi|=0$, $|\mathbf{k}_\xi|=0$. We thereby conclude that conditions \eqref{cphi_condition} and \eqref{cxi_condition} are sufficient for $\tilde{A}_{\varphi\varphi} > 0$ and $\tilde{A}_{\xi\xi}> 0$, if $|\mathbf{k}_\varphi|< k_\varphi^{+}$ and $|\mathbf{k}_\xi|< k_\xi^{+}$. On the other hand, there is a topological lower bound on the admissible values of $k_\varphi$, $k_\xi$ due to the Poincar\'e inequality, 
\begin{eqnarray}
&& \int_Dd^2x\,|\mathbf{k}_\varphi|^2(\delta\varphi)^2 \\
&& \hspace{1cm} = 
\int_Dd^2x\,|\nb\delta\varphi|^2\geq C^{-1}\int_Dd^2x\, (\delta\varphi)^2\,,
\nn
\end{eqnarray}
i.e.,  $\langle \left(|\mathbf{k}_\varphi|^2-C^{-1}\right)(\delta x)^2 \rangle \geq 0$ where $x=\varphi,\xi$. Here $C$ is the Poincar\'e constant depending on the geometry of the domain $D$. Note that for smooth and bounded domains, the smallest eigenvalue of the Laplacian is an optimal value for $C^{-1}$ since it minimizes the Rayleigh quotient.  Finally, note that if we do not assume $\delta\rho=0$, then an additional inequality of the form $\tilde{A}_{\varphi\varphi}\tilde{A}_{\xi\xi}\tilde{A}_{\rho\rho}-\tilde{A}_{\varphi\varphi}\tilde{A}_{\xi\rho}^2-\tilde{A}_{\xi\xi}\tilde{A}_{\varphi\rho}^2> 0$ will emerge. In this case it turns out that $|\bv|^2+d_e^2|\bJ|^2/\rho^2< c_s^2$ is again necessary but not sufficient for stability. Possibly, similar manipulations to those employed above to arrive at sufficient conditions could be used; such a treatment though, would introduce additional constraints on the admissible equilibria and the values of $|\mathbf{k}_x|$, restricting the range of applicability of the resulting stability criterion, which will diverge even more from necessity. %For this reason this analysis will not be pursued. Considering incompressible perturbations ($\delta\rho=0$), which are considered to be the most dangerous, the stability matrix is a $2\times 2$ diagonal matrix with diagonal elements given by $\tilde{\Ac}_{\varphi\varphi}$ and $\tilde{\Ac}_{\xi\xi}$, thus leading to the following sufficient conditional stability criterion 
Summarizing,  the following sufficient conditional stability criterion holds for incompressible perturbations 
%%%%%%%%%
\begin{eqnarray}
& c_\varphi > 0\,,\quad c_\xi > 0\,, \nn\\
& \textrm{for}\quad |\mathbf{k}_\varphi|< k_\varphi^{+}\,,\quad |\mathbf{k}_\xi|< k_\xi^{+}\,, \nn\\
& \langle \left(|\mathbf{k}_x|^2-C^{-1}\right)(\delta x)^2 \rangle \geq 0\,, \label{cond_stab_II}
\end{eqnarray}
where
%%%%%%
\begin{eqnarray}
k_x^+=\frac{1}{2a_x}(-b_x-\sqrt{b_x^2-4 a_x c_x})\,,\quad x=\varphi,\xi\,.
\end{eqnarray}
%%%%%%%%%
Note that the last inequality in \eqref{cond_stab_II} is satisfied for sure if $min(|\mathbf{k}_x|^2)\geq C^{-1}$ and hence, $c_x > 0\,,\; x=\varphi,\xi$, are sufficient stability conditions if $C^{-1}\leq|\mathbf{k}_x|^2< k_x^+$. As a final point we stress that this stability criterion is general enough to capture a large variety of modes as long as $k^+$'s are large enough. Hence, this criterion is practically useful to assess the stability properties of equilibria, when the equilibrium states under consideration render $k^+$'s as large as possible. %Then the validity of the criterion would ensure stability of modes with wavelengths shorter than the $k^+$'s.

%%%%%%%%%%%%%%%%%%%%%%%%%%%%%
%%%%%%%%%%%%%%%%%%%%%%%%%%%%%
%%%%%%%%%%%%%%%%%%%%%%%%%%%%%

\section{Dynamically accessible variations} 
\label{sec_III}

Within the noncanonical Hamiltonian framework, one can consider also the so-called dynamically accessible variations (DAVs) introduced in  \cite{Morrison1989,Morrison1990,Morrison1998}) and used in the MHD context in \cite{Hameiri2003,Andreussi2013,Andreussi2016}. The EC method is valid for general perturbations, but applicable only for EC equilibria and as mentioned in the previous section, many times the perturbations need be restricted to be spatially symmetric. On the other hand, this defect is removed for DA stability analyses, which allow one to treat  generic equilibria by restricting  perturbations to adhere to  phase space constraints; i.e.,  perturbations  are restricted to lie on  the symplectic leaves, which are essentially the level sets of the Casimirs.  Because DAVs  lie on the symplectic leaves they conserve the Casimirs,  that is,  $\delta\Cc_{da}=0$, regardless of the equilibrium conditions.   \\ \indent
In \cite{Morrison1989,Morrison1990,Morrison1998}) it was argued that stability under DAVs  is important because perturbations away from the symplectic leaf of the equilibrium under consideration, although well posed as an initial value problem,  must come from physics outside the dynamical model being considered, since that dynamics preserves the Casimirs.  If such physics is operative,  then one might need to incorporate it into the dynamical model under consideration.  If this were done, then EC or any other kind of stability analysis would likely change.  Viewed this way,  DA stability is quite natural to consider. \\ \indent
In addition to satisfying $\delta\Cc_{da}=0$,  the first order DAVs nullify the Hamiltonian on generic equilibrium points, including the energy-Casimir ones; thus,
\begin{equation}
\delta\Hc[\mathbf{u}_e;\delta \mathbf{u}_{da}]=0\,
\end{equation} 
is a variational principle for generic equilibria. The sufficient stability criterion is provided by the positive definiteness of perturbation energy 
\begin{eqnarray}
\delta^2\Hc_{da}[\mathbf{u}_e]&=& 
\int d^3x\,\Bigg(\frac{\delta^2\Hc}{\delta u^i\delta u^j}\bigg|_{\mathbf{u}_e}\delta u^i_{da}\delta u^j_{da}
\nn\\
&&\hspace{2cm} +\frac{\delta \Hc}{\delta u^i}\bigg|_{\mathbf{u}_e}\delta^2u^i_{da}\Bigg)\,, \label{d2H_da-gen}
\end{eqnarray}
%This particular form of variations was firstly introduced by Arnold \cite{Arnold1965} who used them to study hydrodynamic stability under %what he called ``isovortical'' variations, and later on by Isichenko \cite{Isichenko1998} for MHD, using approaches different from the 
%more efficient and rigorous methodology of producing these perturbations using the Poisson bracket \cite{Morrison1998,Hameiri2003}, %which is the appropriate structure to 
where $\delta u_{da}$ and $\delta^2u_{da}$ are, respectively, first order and second order projections of  arbitrary variations onto the symplectic leaves.  Such DAVs are obtained from the generating functional given by $\Wc=\int d^3x\, u_i\mathrm{g}^i$, where $\bg$ is a state vector embodying the arbitrariness of the perturbations of the various dynamical variables.  The DAVs to first order are given by $\delta\mathbf{u}_{da}=\{\mathbf{u},\Wc\}$. In our case one has
%%%
\begin{eqnarray}
\Wc=\int_Vd^3x\,\left(\mathrm{g}_0\rho+\bg_1\cdot\bv+\bg_2\cdot\bB^*\right)\,, \label{generator}
\end{eqnarray}  
%%%%%%%%%%%%%%%
generating the following variations:
%%%%%%%%
\begin{eqnarray}
\delta\rho_{da}&&=\{\rho,\Wc\}=-\nb\cdot\bg_1\,,\label{drho_da}\\
\delta\bv_{da}&&=\{\bv,\Wc\}\nn\\
&&=-\nb \mathrm{g}_0+\rho^{-1}\bg_1\times\bom+\rho^{-1}(\nb\times\bg_2)\times\bB^*\,,\label{dv_da}\\
\delta\bB^*_{da}&&=\{\bB^*,\Wc\}=\nb\times\big[\rho^{-1}(\bg_1-d_i\nb\times\bg_2)\times\bB^*\nn\\
&&\hspace{2cm}+d_e^2\rho^{-1}(\nb\times\bg_2)\times\bom\big]\,.\label{dB_da}
\end{eqnarray}
To show that the dynamically accessible variation of the Hamiltonian vanishes at general equilibria, we consider
%%%%%%%%%%
\begin{eqnarray}
\delta \Hc_{da}&=&\int_Vd^3x\,\bigg[\rho\bv\cdot\delta\bv_{da}\label{dH_da}\\
&&+\left(h+\frac{v^2}{2}+d_e^2\frac{|\bJ|^2}{2\rho}\right)\delta\rho_{da}+\bB\cdot\delta\bB^*_{da}\bigg]\,,\nn
\end{eqnarray}
%%%%%%%%%%%%
with expressions \eqref{drho_da}--\eqref{dB_da}.  Upon  performing integrations by parts  and omitting the surface integrals, we find  
%%%%%%%%%%
\begin{eqnarray}
\delta \Hc_{da}&=&-\int_Vd^3x\,\bigg\{-\mathrm{g}_0\nb\cdot(\rho\bv)
\label{dH_da}\\
&+& \bg_1\cdot\bigg[\bv\times\bom-\nb\left(h+\frac{v^2}{2}+d_e^2\frac{|\bJ|^2}{2\rho}\right)+\frac{\bJ\times\bB^*}{\rho}\bigg]\nn\\
&+& \bg_2\cdot\nb\times\left[\bv\times\bB^*-d_i\frac{\bJ\times\bB^*}{\rho}+d_e^2\frac{\bJ\times\bom}{\rho}\right]\bigg\}\,.\nn
\end{eqnarray}
%%%%%%%%%%%%
It is apparent that the coefficients of $g_0,\bg_1,\bg_2$ vanish in view of generic XMHD equilibrium conditions and consequently $\delta\Hc_{da}[\mathbf{u_e}]=0$.\\ \indent
To proceed with the derivation of stability criteria,  we need to calculate the second order variation of the Hamiltonian, which in view of Eq.~\eqref{d2H_da-gen}, is 
%%%%
\begin{eqnarray}
&&\delta^2\Hc_{da}=\int_Vd^3x\,\bigg\{\rho|\delta\bv_{da}|^2+\left(h+\frac{v^2}{2}+d_e^2\frac{|\bJ|^2}{2\rho^2}\right)\delta^2\rho_{da}\nn\\
&&+\left[h'(\rho)-d_e^2\frac{|\bJ|^2}{\rho^3}\right](\delta\rho_{da})^2+2\bv\cdot\delta\bv_{da}\delta\rho_{da}+\rho\bv\cdot\delta^2\bv_{da}\nn\\
&&+\delta\bB_{da}\cdot\delta\bB^*_{da}+\bB\cdot\delta^2\bB^*_{da}+\frac{d_e^2}{\rho^2}\bJ\cdot\delta\bJ_{da}\delta\rho_{da}
\bigg\}\,.\label{d2H_da}
\end{eqnarray}
%%%%
From the definition of $\bB^*$ one has
%%%%%%%
\begin{eqnarray}
\delta \bB^*_{da}=&&\delta\bB_{da}\nn\\
&&-d_e^2\nb\times \left(\frac{\bJ}{\rho^2}\delta\rho_{da}\right)+d_e^2\nb\times\left(\frac{\delta\bJ_{da}}{\rho}\right)\,. \label{dB*}
\end{eqnarray}
%%%%%%%%
Upon inserting \eqref{dB*} into \eqref{d2H_da}, the second term of \eqref{dB*} cancels out the last term in \eqref{d2H_da}, leading to
%%%%%%%5 
\begin{eqnarray}
&&\delta^2\Hc_{da}=\int_Vd^3x\,\bigg\{\rho\big|\delta\bv_{da}+\rho^{-1}\bv\delta\rho_{da}\big|^2+|\delta\bB_{da}|^2\nn\\
&&+d_e^2\frac{|\delta\bJ_{da}|^2}{\rho}+\rho^{-1}\left(c_s^2-|\bv|^2-d_e^2\frac{|\bJ|^2}{\rho^2}\right)(\delta\rho_{da})^2
\nn\\
&& +\rho\bv\cdot\delta^2\bv_{da} +\bB\cdot\delta^2\bB^*_{da} \nn\\
&&+\left(h+\frac{|\bv|^2}{2}+d_e^2\frac{|\bJ|^2}{2\rho^2}\right)\delta^2\rho_{da}\bigg\}\,.
\end{eqnarray}
The second order variations of the field variables are given by 
%%%
\begin{widetext}
\begin{eqnarray}
\delta^2\rho_{da}&&=0, \label{d^2rho_da}\\
\delta^2\bv_{da}&&=\rho^{-1}\bg_1\times\nb\times\delta\bv_{da}+\rho^{-1}(\nb\times\bg_2)\times\delta\bB^*_{da}-\rho^{-2}[\bg_1\times\bom+(\nb\times\bg_2)\times\bB^*]\delta\rho_{da}\nn\\
&&=\rho^{-1}(\bzt\times\bom+\bht\times
\bB^*)\nb\cdot(\rho\bzt)+\bzt\times\nb\times\left(\bzt\times\bom+\bht\times\bB^*\right)+\bht\times\nb\times\left[(\bzt-d_i\bht)\times\bB^*+d_e^2\bht\times\bom\right]\,,\label{d^2v_da}\\
\delta^2\bB^*_{da}&&=\nb\times\big\{\rho^{-1}(\bg_1-d_i\nb\times\bg_2)\times\delta \bB^*_{da}+d_e^2\rho^{-1}(\nb\times \bg_2)\times\nb\times\delta\bv_{da}\nn\\
&&-\rho^{-2}[(\bg_1-d_i\nb\times\bg_2)\times\bB^*+d_e^2(\nb\times\bg_2)\times\bom]\delta\rho_{da}\big\}=\nb\times\big\{(\bzt-d_i\bht)\times\nb\times\left[(\bzt-d_i\bht)\times\bB^*+d_e^2\bht\times\bom\right]\nn\\
&&+d_e^2\bht\times\nb\times\left(\bzt\times\bom+\bht\times\bB^*\right)+\rho^{-1}[(\bzt-d_i\bht)\times\bB^*+d_e^2\bht\times\bom]\nb\cdot(\rho\bzt)\big\}\,,\label{d^2B_da}
%\\
%\delta^2\bB^*_{da}&&=\nb\times\left\{\bzt\times\delta\bB^*_{da}-d_i\bht\times\delta \bB^*_{da}+d_e^2\bht\times\nb\times\delta\bv_{da}-\rho^{-2}[(\bzt-d_i\bht)\times\bB^*+d_e^2\bht\times\bom]\delta\rho_{da}\right\}\nn\\
%&&=\nb\times\big\{(\bzt-d_i\bht)\times\nb\times\left(\bzt\times\bB^*-d_i\bht\times\bB^*+d_e^2\bht\times\bom\right)+d_e^2\bht\times\nb\times\left(\bzt\times\bom+\bht\times\bB^*\right)\nn\\
%&&+\rho^{-1}[(\bzt-d_i\bht)\times\bB^*+d_e^2\bht\times\bom]\nb\cdot(\rho\bzt)\big\}\,,\label{d^2B_da}
\end{eqnarray}
\end{widetext}
where $\bzt:=\rho^{-1}\bg_1$ and $\bht:=\rho^{-1}\nb\times \bg_2$  have been  introduced to facilitate the comparison with previous MHD and HMHD results \cite{Hameiri2003,Andreussi2013,Hirota2006}. Evidently,  $\nb\cdot(\rho\bht)=0$ holds by definition of $\bht$.
%%%
%A  simple application of the stability condition derived above is provided in \ref{appendix_A}. \\ \indent
Substituting \eqref{d^2rho_da}--\eqref{d^2B_da} into \eqref{d2H_da}, we find via some straightforward calculations an expression for $\delta^2\Hc_{da}$ (see Appendix \ref{appendix_A}) that is  difficult to be compared with the corresponding HMHD and MHD expressions derived in \cite{Hirota2006} and \cite{Hameiri2003}, respectively. However, after some tedious but also straightforward manipulations, \eqref{d2H_da_xmhd} can be brought in the following form:
%%%%%%%%%%
\begin{widetext}
\begin{eqnarray}
&&\delta^2\Hc_{da}=\int_Vd^3x\,\bigg\{\rho\big|-\nb\mathrm{g}_0+\bzt\times\bom+\bht\times\bB^*+\bzt\cdot\nb\bv-\bv\cdot\nb\bzt\big|^2+\big|\delta\bB_{da}\big|^2-\rho\bzt\cdot\nb[h'(\rho)\nb\cdot(\rho\bzt)]\nn\\
&&-(\bzt\cdot\nb h)\nb\cdot(\rho\bzt)-\bzt\cdot(\bv\cdot\nb\bv)\nb\cdot(\rho\bzt)-(\bzt\times\bJ)\cdot\nb\times(\bzt\times\bB^*) -\rho\bzt\cdot\left[(\bzt\cdot\nb\bv-\bv\cdot\nb\bzt)\cdot\nb\bv\right]\nn\\
&&-\rho\bzt\cdot(\bv\cdot\nb)(\bzt\cdot\nb\bv-\bv\cdot\nb\bzt)+2d_i(\bzt\times\bJ)\cdot\nb\times(\bht\times\bB^*)-d_i\rho(\bht\times\bB^*)\cdot\left[\bht\cdot\nb(\bv-d_i\bJ/\rho)-(\bv-d_i\bJ/\rho)\cdot\nb\bht\right]\nn\\
&&+d_e^2\rho^{-1}\big|\nb\times\delta\bB_{da}\big|^2-d_e^2\bzt\cdot\nb\left(\frac{|\bJ|^2}{2\rho^2}\right)\nb\cdot(\rho\bzt)+d_e^2\rho\bzt\cdot\nb\left[\frac{|\bJ|^2}{\rho^3}\nb\cdot(\rho\bzt)\right]-d_e^2(\bht\times\bJ)\cdot\nb\times(\bht\times\bB^*)\nn\\
&&-d_e^2\left[(2\bzt-d_i\bht)\times\bJ\right]\cdot\nb\times(\bht\times\bom)-d_e^2\rho (\bht\times\bv)\cdot\nb\times(\bht\times\bom)\bigg\}\,. \label{d2H_da_xmhd_2}
\end{eqnarray}
\end{widetext}
%%%%%%%%%%%
Now, it  becomes clear that the case $d_e=0$ corresponds to the barotropic counterpart of the HMHD $\delta^2\Hc_{da}$ given in \cite{Hirota2006}, while if we further impose $d_i=0$ we find $\delta^2\Hc_{da}=\int_Vd^3x\,\rho \big |\delta\bv_{da}+\bzt\cdot\nb\bv-\bv\cdot\nb\bzt|^2+\delta W$ where $\delta W$ is the Frieman-Rotenberg expression for the potential energy \cite{Frieman1960},   consistent with the results found in \cite{Hameiri2003,Andreussi2013}.\\ \indent
The correct MHD limit of \eqref{d2H_da_xmhd_2} reveals an important advantage of the DA method compared to the EC one. As it has been highlighted in \cite{Kaltsas2017,Yoshida2013,Hameiri2013},  the MHD limit of the Casimirs and variational functionals (e.g. the Lagrangian) of XMHD and HMHD, presents certain peculiarities because the Hall term gives rise to singular perturbations, making the derivation of their MHD counterparts rather not straightforward, a difficulty that, as regards to the Casimirs, was treated in \cite{Kaltsas2017} and \cite{Hameiri2013}. Hence, it is natural that this complication is inherited by the  variational principles involving the Casimirs, e.g.,  the EC method. However, in the derivation of $\delta^2\Hc_{da}$ we did not make use of the Casimirs, and therefore their problematic MHD limit does not affect the MHD limit of the DA stability criterion. \\ \indent
The Dirichlet stability theorem, the condition $\delta^2\Hc_{da}>0$ $\forall\; \bzt,\,\bht,\,\mathrm{g}_0$,  ensures the stability of generic XMHD equilibria under dynamically accessible perturbations. However, as long as the variation of the magnetic field is treated as arbitrary, i.e., independent of $\bzt$ and $\bht$, even though it is not, the criterion is based on the positiveness of the terms that do not contain $\delta\bB_{da}$. Thus, we understand that an improvement of the stability criterion can be obtained upon relating $\delta\bB_{da}$ with $\bzt$ and $\bht$ by solving the differential equation that connects $\delta\bB_{da}$ with $\delta\bB^*_{da}(\bzt,\bht)$ and $\delta\rho_{da}(\bzt)$ and follows from the definition of $\bB^*$.
%%%%%%%%%%
%\begin{eqnarray}
%&&\delta\bB_{da}+d_e^2\nb\times\left(\frac{\nb\times\delta\bB_{da}}{\rho_e}\right)=\nb\times\nn\\
%&&\times\left[(\bzt-d_i\bht)\times\bB^*_e+d_e^2\bht\times\bom_e-d_e^2\frac{\bJ_e}{\rho_e^2}\nb\cdot(\rho_e\bzt)\right]\,,
%\end{eqnarray}
%%%%%%%%%
The solution can be effected by introducing a tensorial Green's function as follows:
%%%%%%%
\begin{align}
\delta\bB_{da}=\int_{V'}d^3x'\,\mathbf{G}(\mathbf{x}',\mathbf{x})\cdot\nb\times\nn\\
\left[(\bzt-d_i\bht)\times\bB^*+d_e^2\bht\times\bom-d_e^2\frac{\bJ}{\rho^2}\nb\cdot(\rho\bzt)\right]\,, \label{Green_1}
\end{align}
with $\mathbf{G}(\mathbf{x}',\mathbf{x})$ being the solution of 
%%%%%%%
\begin{eqnarray}
\left[1+d_e^2\nb\times\left(\frac{\nb\times}{\rho}\right)\right]\mathbf{G}_i(\mathbf{x}',\mathbf{x})=\boldsymbol{e}_i\delta(\mathbf{x}'-\mathbf{x})\,,  \label{Green_2}
\end{eqnarray}
%%%%%%%%
with $ i=1,2,3$. For $\rho=const.$ things are simpler since the operator on the lhs of  \eqref{Green_2} becomes the Helmholtz operator (because $\nb\cdot\delta\bB_{da}=0$) and if cartesian coordinates are employed then the equation splits into a set of three independent differential equations, one for each spatial component, in which case  the Green's tensor can be replaced by a scalar Green's function that can be written as an infinite sum of Helmholtz basis functions. The problem, though,  remains  highly dependent on the particular boundary conditions.
%%
%%%%%%%%%%%%%%%%%%%%%%%%%%%%%
%%%%%%%%%%%%%%%%%%%%%%%%%%%%%
%%%%%%%%%%%%%%%%%%%%%%%%%%%%%

\section{Perturbations in mixed Eulerian-Lagrangian framework}
\label{sec_IV}
\begin{figure}
\includegraphics[scale=0.5]{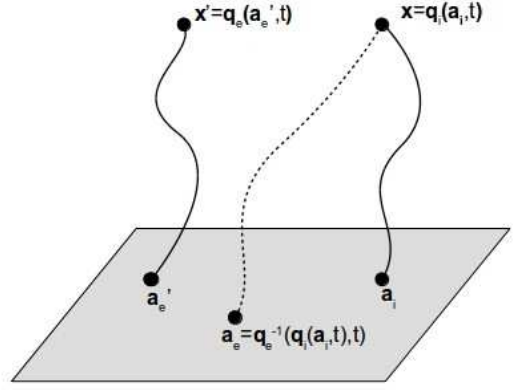}
\caption{The trajectories of a random pair of electron and ion fluid elements labeled by $\ba_e'$ and $\ba_i$, respectively, end up at different locations at time $t>0$. However, if the electron label is chosen so that  $\ba_e=\bq_e^{-1}(\bq_i(\ba_i,t))$ then the trajectories intersect at time $t>0$. \label{fig_2}}
\end{figure}

%\label{sec_II}
%
In the Lagrangian framework, the fluids are not described in terms of fields  measured at fixed position $\mathbf{x}\in \mathbb{R}^3$ 
as in the Eulerian framework adopted above, but in terms of Lagrangian or material variables suitable for tracking the motion of the individual fluid elements. The material variables are the positions of the fluid elements at given instant: $\mathbf{q}_s(\mathbf{a}_s,t)$ ($s=i,e$ standing for the ion and electron species) where $\mathbf{a}_s\in \mathbb{R}^3$ are the fluid element labels, usually taken as the element's position at $t=0$. The two viewpoints are connected through the so-called Lagrange-Euler map, which has to be consistent in the sense that an action written in the Lagrangian framework is mapped to an action written exclusively in terms of Eulerian variables, a requirement called the Eulerian Closure Principle (ECP) \cite{Morrison2009,Morrison2014}. For a two-fluid theory, which is the starting point of the XMHD model, the Lagrange-Euler map is described by the following relations 
%%%%%%%%%%%%%%%%%%
\begin{eqnarray}
\bv_s(\bx,t)=\dot{\bq}_s(\ba_s,t)\bigg|_{\ba_s=\bq_s^{-1}(\bx,t)}\,, \label{L-E_map_1}\\
n_s(\bx,t)=\frac{n_{s0}(\ba_s)}{\Jc_s(\ba_s,t)}\bigg|_{\ba_s=\bq_s^{-1}(\bx,t)}\,, \label{L-E_map_2}\\
s_s(\bx,t)=s_{s0}(\ba_s)\big|_{\ba_s=\bq_s^{-1}(\bx,t)}\,, \label{L-E_map_3}
\end{eqnarray}
%%%%%%%%%%%%%%%%%%
where $s_s$ are the specific entropies of the fluids and $\Jc_s$ $(s=i,e)$, are the Jacobians of $\bq_s$ with respect to $\ba_s$, i.e. $\Jc_s:=det(\partial q_s^i/\partial a_s^j)$. For barotropic fluids, $s_s$ are just constants. Equations \eqref{L-E_map_1}--\eqref{L-E_map_3} are nothing more than the well known single fluid Lagrange-Euler map, described in detail in \cite{Morrison1998}, written for each one of the constituent fluids. The difference between the single-fluid MHD and the two-fluid case is that in the former model the magnetic field can be expressed in terms of Lagrangian variables, due to the frozen-in property of the magnetic field lines. In the case of HMHD and XMHD one can find similar frozen-in properties \cite{Avignon2016,Lingam2016} as well. However, in XMHD this property concerns generalized magnetic-vorticity fields and as a result only the field $\bB^*$ can be explicitly expressed in terms of the Lagrangian variables. This means that similar expressions for $\bB$ can be found only implicitly through a relation similar to \eqref{Green_1}. This makes a fully Lagrangian description of the XMHD model more involved and less universal than the corresponding description for  MHD, since it requires the solution of a differential equation for $\bB$, which depends on the specific boundary conditions. Another peculiarity is that in a fully Lagrangian description the usual Legendre transform cannot be performed and therefore one need to start with a phase-space Lagrangian \cite{Avignon2016}. One way to get rid of those peculiarities is to sacrifice some information about the relationship of the magnetic field with the fluid motion, describing the former as an independent Eulerian variable. Despite this compromise, the resulting mixed Eulerian-Lagrangian description \cite{Charidakos2014}, is still sufficient in order to perform stability analyses and make comparisons with other stability methods. \\ \indent
Lagrangian stability, being applicable for all possible equilibria and also considering perturbations that are not dynamically restricted or constrained by spatial symmetry, most times appears to be the most generic method available. To perform a stability analysis in terms of Lagrangian displacements, within a fully Lagrangian framework, as in the work of Newcomb \cite{Newcomb1962} for MHD or a mixed Eulerian-Lagrangian framework as was done by  Vuilemin \cite{Vuilemin1965} for the complete two-fluid model (without quasineutrality), we need to start with the Lagrangian of the model and compute its second order variation induced by small perturbations. The two-fluid Lagrangian with Maxwell's term being neglected in view of the assumption $v_A\ll c$ ($v_A$ and $c$ are the Alfv\'en speed and the speed of light, respectively)  \cite{Charidakos2014} is
%Lagrangian of the XMHD model is obtained upon normalizing the complete two-fluid Lagrangian and assuming that $v_A/c\ll 1$, where $v_A$ is the Alfv\'en speed and $c$ is the speed of light. This leads to 
%%%%%%%%%%
%\begin{widetext}
\begin{eqnarray}
\Lc&=&\sum_{s=i,e}\int d^3 \am_s \bigg\{ \frac{1}{2}m_sn_{s0}(\ba_s)\big|\dot{\bq}_s(\ba_s,t)\big|^2 
\nn\\
&&-m_s n_{s0}(\ba_s)U_s\left(s_s,\frac{m_sn_{s0}(\ba_s)}{\Jc_s(\ba_s,t)}\right)\nn\\
&&
+\int d^3x \,\delta(\bx-\bq_s(\ba_s,t))e_sn_{s0}(\ba_s)
\nn\\
&&\hspace{1cm}\times \big[\dot{\bq}_s(\ba_s,t)\cdot\bA(\bx,t) -\Phi(\bx,t)\big]\bigg\}
\nn\\
%&&+\int d^3\am_e \bigg\{ \frac{1}{2}m_en_0(\ba_e)\big|\dot{\bq}_e(\ba_e,t)\big|^2-m_e n_0(\ba_e)U_e\left(s_e,\frac{m_en_0(\ba_e)}{\Jc_e(\ba_e,t)}\right)\nn\\
%&&-\int d^3x \,\delta(\bx-\bq_e(\ba_e,t))en_0(\ba_e)\left[\dot{\bq}_e(\ba_e,t)\cdot\bA(\bx,t)-\Phi(\bx,t)\right]\bigg\}
&& -\frac{1}{2\mu_0}\int d^3x \big|\nb\times\bA(\bx,t)\big|^2\,, \label{quasi_2F_Lagrangian}
\end{eqnarray} 
%\end{widetext}
%%%%%%%%%%%
where $\bA$ and $\Phi$ are the vector and electrostatic potentials, respectively. %Let us impose Lagrangian particle density homogeneity, i.e.,  the assumption that any fluid element belonging  to either the ion or to the electron fluid, contains the same number of particles, that is $n_{i0}(\ba_i)=n_{e0}(\ba_e)=n_0=constant$. In view of \eqref{L-E_map_2}, the assumption of Lagrangian homogeneity along with the imposition of Eulerian quasi-neutrality $n_i(\bx,t)=n_e(\bx,t)$ leads to
%\[
%\Jc_i\big|_{\ba_i=\bq_i^{-1}(\bx,t)}=\Jc_e\big|_{\ba_e=\bq_e^{-1}(\bx,t)}\,.
%\]
 Now, since the trajectories $\bq_i$, $\bq_e$ of the ion and electron fluid elements are in general different,  at time $t>0$ they will be located at different positions $\bx$ and  $\bx'$ unless the fluid elements $\ba_i$ and $\ba_e$ are chosen appropriately so to make $\bx'=\bx$. %Therefore, in general, if we try to write down a single fluid version of \eqref{quasi_2F_Lagrangian} and then take the Lagrange-Euler map, we will end up with a nonlocal Lagrangian density in the Eulerian description. Hence, 
 Imposing locality on  the Eulerian level, is equivalent to matching up the ion and electron fluid elements on the basis of the map $\ba_e=\bq_e^{-1}(\bq_i(\ba_i,t),t)$,  (see Fig. \ref{fig_2} and also the corresponding explanation in \cite{Avignon2016}).
%%%%
%%%%%%%
In view of these considerations we understand that local quasineutrality on the Lagrangian level is ensured if
\begin{eqnarray}
n_{i0}(\ba_i)=n_{e0}(\ba_e)|_{\ba_e=\bq_e^{-1}(\bq_i(\ba_i,t),t)}\,. \label{lagrangian_quasineutrality}
\end{eqnarray}
In view of \eqref{L-E_map_2}, Eq. \eqref{lagrangian_quasineutrality} along with the imposition of Eulerian quasineutrality $n_i(\bx,t)=n_e(\bx,t)$ leads to
\[
\Jc_i\big|_{\ba_i=\bq_i^{-1}(\bx,t)}=\Jc_e\big|_{\ba_e=\bq_e^{-1}(\bx,t)}\,.
\].
 The final step for obtaining an XMHD action is to replace the ion and electron Lagrangian variables with XMHD-like variables, which would play the roles of Lagrangian analogues for $\bv$ and $\bJ/(en)$. In this regard, we define two new Lagrangian quantities $\bQ$ and $\bD$ %that  can be either of local or nonlocal nature. The nonlocal pair of variables can be defined 
 through the following relations:
%%%%%
\begin{eqnarray}
\bQ(\ba_i,\ba_e,t) &:=&
\frac{m_i}{m}\bq_i(\ba_i,t)+\frac{m_e}{m}\bq_e(\ba_e,t)\,,\label{material_xmhd_vars_1}\\
\bD(\ba_i,\ba_e,t)&:=&\bq_i(\ba_i,t)-\bq_e(\ba_e,t)\,. \label{material_xmhd_vars_2}
\end{eqnarray}
%%%%%
The inverse transformation reads as follows:
\begin{eqnarray}
\bq_i(\ba_i,t)&=&\bQ(\ba_i,\ba_e,t)%\big|_{\ba_e=\bq_e^{-1}(\bq_i(\ba_i,t),t)}
+\alpha_i\bD(\ba_i,\ba_e,t)
%\big|_{\ba_e=\bq_e^{-1}(\bq_i(\ba_i,t),t)}
\,,\nn \\
\bq_e(\ba_e,t)&=&\bQ(\ba_i,\ba_e,t)%\big|_{\ba_i=\bq_i^{-1}(\bq_e(\ba_e,t),t)}
 +\alpha_e\bD(\ba_i,\ba_e,t)%\big|_{\ba_i=\bq_i^{-1}(\bq_e(\ba_e,t),t)}
\,,\label{inverse_material_xmhd_vars}
\end{eqnarray}
where $\alpha_i=m_e/m$ and $\alpha_e=-m_i/m$. We are now in position to write down an XMHD Lagrangian in $(\bQ,\bD)$ variables as follows
%\begin{widetext}
\begin{eqnarray}
\Lc=&&\int \int d^3 \am_i d^3 \am_e \delta(\ba_e-\bq_e^{-1}(\bq_i(\ba_i,t),t))\times \nn\\
&&\times \sum_{s=i,e} n_{s0}(\ba_s) \bigg[ \frac{m_s}{2}\big|\dot{\bQ}\big|^2
+\frac{m_s}{2}\alpha_s^2\big|\dot{\bD}\big|^2+m_s\alpha_s\dot{\bQ}\cdot\dot{\bD}\nn\\
&&+e_s\dot{\bQ}\cdot\bA(\bq_s(\ba_s,t),t)-e_s\Phi(\bq_s(\ba_s,t),t)\nn\\
&& +e_s\alpha_s \dot{\bD}\cdot\bA(\bq_s(\ba_s,t),t)-m_s U_s\left(s_s,\frac{m_sn_{s0}(\ba_s)}{\Jc_s(\ba_s,t)}\right)\bigg]\nn\\
&&-\frac{1}{2\mu_0}\int d^3x \big|\nb\times\bA(\bx,t)\big|^2\,. \label{material_xmhd_Lagrangian}
\end{eqnarray} 
%\end{widetext}
Note that $\bQ$ and $\bD$ depend on $\ba_i$ and $\ba_e$. Also, the role of the delta function is to ensure the locality of the Eulerian version of \eqref{material_xmhd_Lagrangian}, i.e., the trajectories $\bq_i$ and $\bq_e$ meet each other at $t>0$. In general, we are interested in examining the stability of stationary equilibria in the Eulerian picture. It is well known \cite{Morrison1998,Newcomb1962,Andreussi2013} that not all Eulerian equilibria correspond to Lagrangian ones, e.g.,  % stationary Eulerian equilibria with $\bv=\bv(\mathbf{x})$ cannot correspond to Lagrangian equilibria, because the latter means $\dot{\bq}=0$ which corresponds to $\bv=0$. This is simple to be understood since 
for an Eulerian equilibrium state with flow, an infinite number of fluid elements have to be in motion for the realization of this flow. However, in the Lagrangian framework, moving fluid elements correspond to time dependent material variables. Therefore, we conclude that stationary Eulerian states correspond to time dependent Lagrangian trajectories $\bq_{s0}=\bq_{s0}(\ba_s,t)$. Hence, we expand the material variables around time dependent reference trajectories considering a small perturbation, that is, the fields should be decomposed as follows
\begin{eqnarray}
\bQ(\ba_i,\ba_e,t)&=&\bQ_0(\ba_i,\ba_e,t)+\bzt(\ba_i,\ba_e,t)\,, \label{perturb_1} \\
\bD(\ba_i,\ba_e,t)&=&\bD_0(\ba_i,\ba_e,t)+\bht(\ba_i,\ba_e,t)\,,\label{perturb_2}\\
\bA(\bx,t)&=&\bA_0(\bx)+\bA_1(\bx,t)\,,\label{perturb_3}\\
\Phi(\bx,t)&=&\Phi_0(\bx)+\Phi_1(\bx,t)\,, \label{perturb_4}
\end{eqnarray}
where the quantities with subscript $0$ define the equilibrium state, those with subscript $1$ define the perturbed electromagnetic field and $\bzt$, $\bht$ are Lagrangian displacements accounting for the perturbation of the fluid element trajectories.  Hence, in view of \eqref{perturb_1}--\eqref{perturb_4}, we  find using  \eqref{material_xmhd_Lagrangian} the  perturbed  $\Lc=\Lc_0+\Lc_1+\Lc_2+\cdots $. For stability   we are interested in $\Lc_2$ because it describes the linearized dynamics, while $\Lc_0$ is merely a constant and $\Lc_1$ vanishes at equilibrium. To write down the second order perturbation of the Lagrangian we need to expand the electromagnetic potentials and the internal energies. The magnetic and electric potentials are computed on the fluid trajectories; thus, up to second order, they are %expanded as follows:
%%%%%%%%%%%%
%\begin{widetext}
\begin{eqnarray}
 &&\bA(\bq_{s0}+\bzt+\alpha_s\bht,t)=\bA_0(\bq_{s0})+\bA_1(\bq_{s0},t)
\label{sec_order_EM_interaction_1}\\
 &&\qquad  +(\bzt+\alpha_s\bht)\cdot \nb_{\bq_{s0}}\bA_0(\bq_{s0})
 \nn\\
 &&\qquad + (\bzt+\alpha_s\bht)\cdot\nb_{\bq_{s0}}\bA_1(\bq_{s0},t)
\nn\\
&&\qquad  +\frac{1}{2}(\bzt+\alpha_s\bht)(\bzt+\alpha_s\bht)\bdd\nb_{\bq_{s0}}\nb_{\bq_{s0}} \bA_0(\bq_{s0})\,,
 \nn\\
&& \Phi(\bq_{s0}+\bzt+\alpha_s\bht,t)=\Phi_0(\bq_{s0})+\Phi_1(\bq_{s0},t)
\label{sec_order_EM_interaction_2}\\
 &&\qquad +(\bzt+\alpha_s\bht)\cdot\nb_{\bq_{s0}}\Phi_0(\bq_{s0})
\nn\\
&&\qquad +(\bzt+\alpha_s\bht)\cdot\nb_{\bq_{s0}}\Phi_1(\bq_{s0},t)
\nn\\
&&\qquad +\frac{1}{2}(\bzt+\alpha_s\bht)(\bzt+\alpha_s\bht)\bdd\nb_{\bq_{s0}}\nb_{\bq_{s0}}\Phi_0(\bq_{s0})\,, \nn
\end{eqnarray}
%\end{widetext}
%%%%%%%%%%%
where $\mathbf{a}\mathbf{b}\bdd\mathbf{c}\mathbf{d}:=a_ib_jc^jd^i$. % and $\nb_{\bq_s}:=\nb_{\bq_{s0}}$. 
The second order perturbative expansion of the internal energy terms is performed along  lines similar to those of   the single fluid case (see \cite{Morrison1998}) in Appendix \ref{appendix_B}. 
%%%%%%%%%%%%%%
Henceforth, the subscript $0$ will be dropped on the understanding that from now on $\bA, \Phi$, $\bQ$, $\bD$, $\bq_s$ and $\Jc_s$ correspond to equilibrium. Using the results \eqref{sec_order_EM_interaction_1}--\eqref{sec_order_EM_interaction_2} and \eqref{sec_order_internal_energy} we are able to construct $\Lc_2$
%%%%%%%%%%%%%%%
\begin{widetext}
\begin{eqnarray}
\Lc_2=&&\int \int d^3 \am_i d^3 \am_e \delta(\ba_e-\bq_e^{-1}(\bq_i(\ba_i,t),t)) \sum_sn_{s0}(\ba_s)\bigg\{ \frac{m_s}{2}\big|\dot{\bzt}\big|^2+\alpha_s^2\frac{m_s}{2}\big|\dot{\bht}\big|^2+m_s\alpha_s\dot{\bzt}\cdot\dot{\bht}\nn\\
&&+e_s\left(\dot{\bQ}+\alpha_s\dot{\bD}\right)\cdot\big[ \left(\bzt+\alpha_s\bht \right)\cdot\nb_{\bq_s}\bA_1(\bq_{s},t)+\frac{1}{2}\left(\bzt+\alpha_s\bht \right)\left(\bzt+\alpha_s\bht \right)\bdd\nb_{\bq_s}\nb_{\bq_s}\bA(\bq_{s})\big]\nn\\
&&+e_s(\dot{\bzt}+\alpha_s\dot{\bht})\cdot\big[\bA_1(\bq_{s},t)+ \left(\bzt+\alpha_s\bht \right)\cdot\nb_{\bq_s}\bA(\bq_{s})\big]-e_s\left(\bzt+\alpha_s\bht \right)\cdot\nb_{\bq_s}\Phi_1(\bq_{s},t)\nn\\
&&-\frac{e_s}{2}\left(\bzt+\alpha_s\bht \right)\left(\bzt+\alpha_s\bht \right)\bdd\nb_{\bq_s}\nb_{\bq_s}\Phi(\bq_{s})-\frac{n_{s0}^2}{2\Jc_{s}^2}\frac{\partial^2\Uc_s}{\partial n_s^2}\left(\nb_{\bq_s}\cdot\bzt+\alpha_s\nb_{\bq_s}\cdot\bht\right)^2\nn\\
&&-\frac{n_{s0}}{2\Jc_{s}}\frac{\partial\Uc_s}{\partial n_s}\left[\left(\nb_{\bq_s}\cdot\bzt+\alpha_s\nb_{\bq_s}\cdot\bht\right)^2+\nb_{\bq_s}\left(\bzt+\alpha_s\bht\right)\bdd\nb_{\bq_s}\left(\bzt+\alpha_s\bht\right)\right]\bigg\} -\frac{1}{2\mu_0}\int d^3x \big|\nb\times\bA_1(\bx,t)\big|^2\,, \label{perturbed_XMHD_Lagrangian_material}
\end{eqnarray}
\end{widetext}
%%%%%%%%%%%%%%
where $\Uc_s=m_sU_s$. The locality of the perturbed Lagrangian density is imposed through the delta function in \eqref{perturbed_XMHD_Lagrangian_material} by means of the equilibrium trajectories, i.e., $\delta(\ba_e-\bq_{e0}^{-1}(\bq_{i0}(\ba_i,t),t))$. This is equivalent to imposing $\bx'=\bx$, after performing a single fluid Lagrange-Euler map (which involves the unperturbed trajectories, e.g. \cite{Morrison1998}) for each constituent fluid. The Lagrangian \eqref{perturbed_XMHD_Lagrangian_material} is not very different from the two-fluid result of Vuilemin \cite{Vuilemin1965}; actually,  it is the quasineutral counterpart of his second order perturbed Lagrangian, written however in terms of the XMHD Lagrangian displacements $\bzt$, $\bht$ instead of the two-fluid ones $\boldsymbol{\xi}_i$, $\boldsymbol{\xi}_e$.  Moreover, \eqref{perturbed_XMHD_Lagrangian_material} is applicable for generic thermodynamic closures with scalar pressure, not only for fluids obeying the adiabatic ideal-gas law as in \cite{Vuilemin1965}. The most important advantage of our formulation can be seen though, after employing the Lagrange-Euler map:  first because \eqref{perturbed_XMHD_Lagrangian_material} explicitly dictates how the labels of the fluid elements are related so that the Lagrange-Euler map will result in a local Lagrangian and second because its Eulerian counterpart will be expressed in terms of the MHD-like variables  $\bv$ and $\bJ$.\\ \indent
To employ the Lagrange-Euler map we need to ``Eulerianize'' the displacement vectors. This procedure, along with the calculation of the Eulerian-field variations in terms of the Lagrangian displacements, which enables us to compare them with DAVs, is presented in Appendix \ref{appendix_C}. The Eulerian variations of the fields are
\begin{widetext}
\begin{eqnarray}
\delta\bv&=& \partial_t\tbzt+\bv\cdot\nb\tbzt-\tbzt\cdot\nb\bv+\frac{m_im_e}{m^2}\left(\bw\cdot\nb\tbht-\tbht\cdot\nb\bw\right) \,, \label{delta_v}\\
\delta\bw&=&\partial_t\tbht+\bv\cdot\nb\tbht-\tbht\cdot\nb\bv+\bw\cdot\nb\tbzt-\tbzt\cdot\nb\bw+\frac{m_e^2-m_i^2}{m^2}\left(\bw\cdot\nb \tbht-\tbht\cdot\nb\bw\right) \,, \label{delta_w} 
\end{eqnarray}
where $\bw:=\bJ/(en)$ and $\tbzt$, $\tbht$ are the Eulerianized displacement vectors.
%%%%%%%%%%%%%%%%
Using the maps \eqref{inverse_dot_Q_D} and \eqref{time_der_euler_displacents} from Appendix \ref{appendix_C}, and also the relations \eqref{L-E_map_2}, \eqref{lagrangian_quasineutrality} together with $d^3x=\Jc_sd^3\am_s$, we  compute the Eulerian expression for $\Lc_2$ from the Lagrangian \eqref{perturbed_XMHD_Lagrangian_material} 
\begin{eqnarray}
\Lc_2=&&\int d^3x\, \bigg\{\frac{mn}{2}\big|\partial_t\bzt\big|^2+\frac{m_im_e}{2m}n\big|\partial_t\bht\big|^2+\partial_t\bzt\cdot\left[mn\left(\bv\cdot\nabla\bzt+\frac{m_im_e}{m^2}\bw\cdot\nb\bht\right)+en\bht\cdot\nb\bA\right]\nn\\
&&+\partial_t\bht\cdot\left[\frac{m_im_e}{m}n\left(\bv\cdot\nb\bht+\bw\cdot\nb\bzt+\frac{m_e^2-m_i^2}{m^2}\bw\cdot\nb\bht\right)+en\left(\bA_1+\bzt\cdot\nb\bA+\frac{m_e^2-m_i^2}{m^2}\bht\cdot\nb\bA\right)\right]\nn\\
&&\hspace{2cm}+\mathfrak{W}(\bzt,\bht,\bA_1,\Phi_1)\bigg\}\,,\label{two-fluid_perturbed_Lagrangian}
\end{eqnarray}
%%%%%%%%%
where
%%%%%%%%%%%
\begin{eqnarray}
\mathfrak{W}(\bzt,\bht,\bA_1,\Phi_1)=&&-\frac{1}{2\mu_0}\big|\nb\times\bA_1\big|^2+ \frac{mn}{2}\big|\bv\cdot\nb\bzt+\frac{m_im_e}{m^2}\bw\cdot\nb\bht\big|^2\nn\\
&&+\frac{m_im_e n}{2m}\big|\bv\cdot\nb \bht+\bw\cdot\nb\bzt+\frac{m_e^2-m_i^2}{m^2}\bw\cdot\nb\bht\big|^2+en\left(\bv\cdot\nb\bzt+\frac{m_im_e}{m^2}\bw\cdot\nb\bht\right)\cdot\left(\bht\cdot\nb\bA\right)\nn\\
&&+en \left(\bv\cdot\nb\bht+\bw\cdot\nb\bzt+\frac{m_e^2-m_i^2}{m^2}\bw\cdot\nb\bht\right)\cdot\left(\bA_1+\bzt\cdot\nb\bA+\frac{m_e^2-m_i^2}{m^2}\bht\cdot\nb\bA\right)\nn\\
&&+en\bigg[\bv\cdot(\bht\cdot\nabla\bA_1)+\bw\cdot(\bzt\cdot\nb\bA_1)+\bv\cdot(\bzt\bht\bdd\nb\nb\bA)+\frac{1}{2}\bw\cdot(\bzt\bzt\bdd\nb\nb\bA)+\frac{m_e^2-m_i^2}{m^2}\bw\cdot(\bht\cdot\nb\bA_1)\nn\\
&&+\frac{m_e^2-m_i^2}{m^2}\bw\cdot(\bzt\bht\bdd\nb\nb\bA)+\frac{m_e^2-m_i^2}{2m^2}\bv\cdot(\bht\bht\bdd\nb\nb\bA)+\frac{m_e^3+m_i^3}{2m^3}\bw\cdot(\bht\bht\bdd\nb\nb\bA)\nn\\
&&-\bzt\bht\bdd\nb\nb \Phi-\frac{m_e^2-m_i^2}{2m^2}\bht\bht\bdd\nb\nb\Phi-\bht\cdot\nb\Phi_1\nn\bigg]-\frac{p}{2}\left[\nb\bzt\bdd\nb\bzt-(\nb\cdot\bzt)^2 \right]-\frac{1}{2}n \frac{\partial p}{\partial n}(\nb\cdot\bzt)^2\nn\\
&&-\left[\nb\bzt\bdd\nb\bht-(\nb\cdot\bzt)(\nb\cdot\bht) \right]\left(\frac{m_e}{m}p_i-\frac{m_i}{m}p_e\right)-n\left(\frac{m_e}{m}\frac{\partial p_i}{\partial n}-\frac{m_i}{m}\frac{\partial p_e}{\partial n}\right)(\nb\cdot\bzt)(\nb\cdot\bht)\nn\\
&&-\frac{1}{2}\left[\nb\bht\bdd\nb\bht-(\nb\cdot\bht)^2 \right]\left[\left(\frac{m_e}{m}\right)^2p_i+\left(\frac{m_i}{m}\right)^2p_e\right]-\frac{1}{2}n\left[\left(\frac{m_e}{m}\right)^2\frac{\partial p_i}{\partial n}+\left(\frac{m_i}{m}\right)^2\frac{\partial p_e}{\partial n}\right](\nb\cdot\bht)^2\,.\label{W_xmhd}
\end{eqnarray}
\end{widetext}
Here we have used $p_s=n^2\partial\Uc_s/\partial n$,   Dalton's law $p=p_i+p_e$,  and in addition $n^3\partial^2\Uc_s/\partial n^2=n\partial p_s/\partial n-2p_s$. Also the tildes have been dropped since we are working now in a completely Eulerian framework and there is no need to distinguish from the Lagrangian variables. We should stress here that the version of the XMHD model we use in the previous sections was derived upon expanding the quasineutral two-fluid equations and keeping terms up to zeroth order in $\mu:=m_e/m_i$ in the Alfv\'en normalized equations of motion.  In the derivations of this section we have not performed such an expansion and therefore up to now our results are fully two-fluid with quasi-neutrality. Hence,  they can be used either to describe an ion-electron plasma or a positron-electron plasma, just by replacing the ion mass by the positron mass.\\ \indent
 The Euler-Lagrange equations that correspond to \eqref{two-fluid_perturbed_Lagrangian} are obtained upon minimizing the action 
\begin{align}
 \Sc_2=\int_{t_1}^{t_2} dt \Lc_2\,, \label{action}
\end{align}
  with boundary conditions $\bzt\cdot\hat{\mathrm{n}}=\bht\cdot\hat{\mathrm{n}}=0$, where $\hat{\mathrm{n}}$ is the unit vector normal to the boundary and 
\begin{eqnarray}
  \bzt(\bx,t=t_1)&=&\bzt(\bx,t=t_2)
 \nn \\
  &=&\bht(\bx,t=t_1)=\bht(\bx,t=t_2)=0
  \nn\,.
\end{eqnarray}
 These equations describe the linearized dynamics;  more specifically,  from the $\bzt$-variation one obtains  the linearized momentum equation,  while from $\bht$-variations a generalized Ohm's law occurs. However,  there are two redundant variables, namely $\bA_1$ and $\Phi_1$,  which do not appear in pairs of generalized coordinates and velocities. In some way, we need to express them in terms of the generalized coordinates so as to eliminate this redundancy.  As regards $\Phi_1$ one can express it by selecting a particular gauge. Alternatively, we can compute the respective ``Euler-Lagrange equations'' that  can be used either to eliminate $\Phi_1$ and $\bA_1$ or as side conditions.  Accordingly, extremizing the action with respect to the electromagnetic field variables, we find
\begin{eqnarray}
\delta\Phi_1:\quad && e\nb\cdot(n\bht)=0\,, \label{delta_Phi}\\
\delta\bA_1:   \quad && en\Big[
\partial_t\bht +\bv\cdot\nb\bht-\bht\cdot\nb\bv+\bw\cdot\nb\bzt-\bzt\cdot\nb\bw
\nn\\ 
&& \qquad +\frac{m_e^2-m_i^2}{m^2}
\left(\bw\cdot\nb\bht-\bht\cdot\nb\bw\right)
\Big]
\nn\\
&&\qquad -\frac{\bJ}{n}\nb\cdot(n\bzt)-\bJ_1=0
\label{delta_A}\,,
\end{eqnarray}
where for the derivation of \eqref{delta_A} we  assumed $(\bA_1\times\delta\bA_1)\big|_{\partial D}\cdot\hat{\mathrm{n}}=0$. Equation \eqref{delta_Phi} expresses charge neutrality for the perturbed state. In view of this condition, the term that contains $\Phi_1$ in $\mathfrak{W}$ can be eliminated upon integrating by parts. Also, in principle  \eqref{delta_A} can be used to express $\bA_1$ in terms of $\bzt$ and $\bht$. Combining Eq.~\eqref{delta_A} with \eqref{delta_w}, we find the expression for the Eulerian variation of the particle density to be
%%%%%%%%%%
\begin{eqnarray}
n_1=-\nb\cdot(n\bzt)\,, \label{delta_n}
\end{eqnarray}
%%%%%%%%%%%%
which is of the form of $\delta \rho_{da}$ (see Eq.~\eqref{drho_da}). \\ \indent
To arrive at a sufficient stability condition we need to calculate the Hamiltonian of the linearized dynamics. To this end, the standard procedure of Legendre transforming the Lagrangian \eqref{two-fluid_perturbed_Lagrangian} can be applied. The departing point for performing this transformation is to define the generalized momenta $\bpi_{\zeta}$ and  $\bpi_{\eta}$ as follows:
\begin{widetext}
\begin{eqnarray}
\bpi_{\zeta}&:=&\frac{\delta\Lc_2}{\delta\dot{\bzt}}=mn\left(\partial_t\bzt+\bv\cdot\nb\bzt+\frac{m_im_e}{m^2}\bw\cdot\nb\bht\right)+en \bht\cdot\nb\bA\,,\label{pi_bzt}\\
\bpi_{\eta}&:=&\frac{\delta\Lc_2}{\delta\dot{\bht}}=\frac{m_im_e}{m}n\left(\partial_t\bht+\bv\cdot\nb\bht+\bw\cdot\nb\bzt+\frac{m_e^2-m_i^2}{m^2}\bw\cdot\nb\bht\right)\nn\\
&&\hspace{1cm}  +en\left(\bA_1+\bzt\cdot\nb\bA+\frac{m_e^2-m_i^2}{m^2}\bht\cdot\nb\bA\right)\,. \label{pi_bht}
\end{eqnarray}
%%%%
With \eqref{pi_bzt} and \eqref{pi_bht} at hand, we employ  the usual Legendre transform,  $\Hc_2=\int_Dd^3x\,\left(\bpi_{\zeta}\cdot\partial_t\bzt+\bpi_{\eta}\cdot\partial_t\bht\right)-\Lc_2$, to find
\begin{eqnarray}
\Hc_2=\int_Dd^3x\,\bigg[\frac{1}{2mn}\bigg|\bpi_{\zeta}-mn\left(\bv\cdot\nb\bzt+\frac{m_im_e}{m^2}\bw\cdot\nb\bht\right)-en\bht\cdot\nb\bA\bigg|^2\nn\\
+\frac{m}{2m_im_en}\bigg|\bpi_\eta-\frac{m_im_e}{m}n\left(\bv\cdot\nb\bht+\bw\cdot\nb\bzt+\frac{m_e^2-m_i^2}{m^2}\bw\cdot\nb\bht\right)\nn\\
-en\left(\bA_1+\bzt\cdot\nb\bA+\frac{m_e^2-m_i^2}{m^2}\bht\cdot\nb\bA\right)\bigg|^2-\mathfrak{W}(\bzt,\bht)\bigg]\,. \label{two_fluid_perturbed_hamiltonian}
\end{eqnarray}
\end{widetext}
From \eqref{two_fluid_perturbed_hamiltonian} we deduce that 
\begin{eqnarray}
-\int d^3x\,\mathfrak{W}(\bzt,\bht)\geq 0
\end{eqnarray}
with $\mathfrak{W}(\bzt,\bht)$ given by \eqref{W_xmhd} implies stability.
%
%
%
%%%%%%%%%%%%%%%%%%%%%%%%%%%%%
%%%%%%%%%%%%%%%%%%%%%%%%%%%%%
%%%%%%%%%%%%%%%%%%%%%%%%%%%%%
%
%
\section{Hall MHD}
The HMHD case has an interesting peculiarity: to derive the HMHD perturbed Lagrangian, we assume massless electrons, i.e., $m_e=0$; as a result, $\partial_t \bht$ appears linearly in $\Lc_2$, and therefore the definition of the canonical momentum $\bpi_\eta$ results in a constraint instead of an equation that can be used to express $\partial_t \bht$ in terms of $\bpi_\eta$. But before addressing this peculiarity, we Alfv\'en normalize the HMHD Lagrangian term by term so as to facilitate comparisons with already known results in this framework. The Alfv\'en normalization is effected by
%%%%%%%%%%%%%%%%%
\begin{eqnarray}
&&\bar{n}=n/n_0\,,\quad  \bar{t}=t/\tau_A\,, \quad   \bar{\nb}=\ell\nb\,, 
\label{normalized_quantities}\\
&& 
 \bar{\bB}=\bB/B_0\,, \quad  \bar{\bJ}=\bJ\big/(B_0/\ell \mu_0)\,, \quad  \bar{\bA}=\bA/(\ell B_0)\,, 
\nn\\
&& \bar{\bE}=\bE/(v_AB_0)\,,\ \bar{\Phi}=\Phi/(\ell v_A B_0)\,,
\   \bar{p_s}=p_s\big/(B_0^2/\mu_0)\,, 
\nn
\end{eqnarray}
%%%%%%%%%%%%%%
where $\ell$, $n_0$ and $B_0$ are a reference length, particle density, and magnetic field, respectively; $v_A=B_0/\sqrt{\mu_0m_in_0}$ is the Alfv\'en speed,  and  $\tau_A=\ell/v_A$ is the Alfv\'en time. In order to write the Lagrangian in dimensionless form, we need also to introduce normalized displacements $\bzt$ and $\bht$. Equations \eqref{delta_v} and \eqref{delta_w} suggest that an appropriate normalization is 
%%%%%%%%%%
\begin{eqnarray}
\bar{\bzt}=\bzt/\ell\,,\quad \bar{\bht}=\bht\big/ \sqrt{m_i/\mu_0n_0e^2}=\bht/\lambda_i\,, \label{normalized_displacements}
\end{eqnarray}
%%%%%%%%%%
where $\lambda_i$ is the ion skin depth $(\lambda_i=d_i\ell)$. In view of \eqref{normalized_quantities} and \eqref{normalized_displacements} and setting $m_e=0$, the Lagrangian \eqref{two-fluid_perturbed_Lagrangian} can be brought into the  following dimensionless form,  
\begin{eqnarray}
\Lc_2&=&\int d^3x \,\bigg\{\frac{\rho}{2}\big|\partial_t\bzt\big|^2+\rho\left(\partial_t \bzt\right)\cdot(\bht\cdot\nb\bA+\bv\cdot\nb\bzt)\nn\\
&&+\rho(\partial_t\bht)\cdot\left(\bA_1+\bzt\cdot\nb\bA-d_i\bht\cdot\nb\bA\right)
\nn\\
&& \quad +\mathfrak{W}_{hmhd}(\bzt,\bht,\bA_1)\bigg\}\,, 
\label{hmhd_perturbed_Lagrangian}
\end{eqnarray}
where
\begin{widetext}
\begin{eqnarray}
\mathfrak{W}_{hmhd}=&&\frac{\rho}{2}\big|\bv\cdot\nb\bzt\big|^2+\rho(\bv\cdot\nb\bzt)\cdot(\bht\cdot\nb\bA)\nn\\
&&+\rho\left( \bv\cdot\nb\bht+\rho^{-1}\bJ\cdot\nb\bzt-d_i\rho^{-1}\bJ\cdot\nb\bht \right)\cdot\left(\bA_1+\bzt\cdot\nb\bA-d_i\bht\cdot\nb\bA\right)+\rho \bv\cdot(\bht\cdot\nb\bA_1)\nn\\
&&+\rho\bv\cdot(\bzt\bht\bdd\nb\nb\bA)-\frac{d_i}{2}\rho\bv\cdot(\bht\bht\bdd\nb\nb\bA)+\bJ\cdot(\bzt\cdot\nb\bA_1)-d_i\bJ\cdot(\bht\cdot\nb\bA_1)+\frac{1}{2}\bJ\cdot(\bzt\bzt\bdd\nb\nb\bA)\nn\\
&&-d_i\bJ\cdot(\bzt\bht\bdd\nb\nb\bA)+\frac{d_i^2}{2}\bJ\cdot(\bht\bht\bdd\nb\nb\bA)-\rho\bht\cdot\nb\Phi_1-\rho (\bzt\bht\bdd\nb\nb\Phi)+\frac{d_i}{2}\rho \bht\bht\bdd\nb\nb\Phi\nn\\
&&-\frac{p}{2}[\nb\bzt\bdd\nb\bzt-(\nb\cdot\bzt)^2]-\frac{\rho}{2}c_s^2(\nb\cdot\bzt)^2+d_ip_e[\nb\bzt\bdd\nb\bht-(\nb\cdot\bzt)(\nb\cdot\bht)]+d_i\rho c_{se}^2(\nb\cdot\bzt)(\nb\cdot\bht)\nn\\
&&-\frac{d_i^2}{2}p_e[\nb\bht\bdd\nb\bht-(\nb\cdot\bht)^2]-\frac{d_i^2}{2}\rho c_{se}^2(\nb\cdot\bht)^2-\frac{1}{2}\big|\bB_1\big|^2\,, \label{W_hmhd}
\end{eqnarray}
\end{widetext}
and the bars have been dropped. Note that the term, $\int d^3x\,\rho \bht\cdot\nb \Phi_1$, vanishes in view of \eqref{delta_Phi} and the boundary conditions. In addition,  the perturbation of the velocity field and of the field $\bJ/\rho$ are given by 
\begin{eqnarray}
\delta\bv&=&\partial_t\bzt+\bv\cdot\nb\bzt-\bzt\cdot\nb\bv\,,\label{dv_hmhd}\\
\delta\left(\frac{\bJ}{\rho}\right)&=&\partial_t\bht+\bv\cdot\nb\bht-\bht\cdot\nb\bv
\label{dw_hmhd}\\ &&+\frac{\bJ}{
\rho}\cdot\nb\bzt-\bzt\cdot\nb\frac{\bJ}{\rho} 
-d_i\left(\frac{\bJ}{\rho}\cdot\nb\bht-\bht\cdot\nb\frac{\bJ}{\rho}\right),\nn
\end{eqnarray}
while the generalized momenta $\bpi_{\bzt}$ and $\bpi_{\bht}$ are now computed as follows 
\begin{eqnarray}
\bpi_{\zeta}&&=\frac{\delta \Lc_2}{\delta \dot{\bzt}}=\rho(\partial_t\bzt+\bv\cdot\nb\bzt)+\rho\bht\cdot\nb\bA\,,\label{pi_zeta_hall}\\
\bpi_{\eta}&&=\frac{\delta \Lc_2}{\delta \dot{\bht}}=\rho\left(\bA_1+\bzt\cdot\nb \bA-d_i \bht\cdot\nb\bA \right). \label{pi_eta_hall}
\end{eqnarray}
%%%%%%%%%%%%
Note that Eq.~\eqref{pi_eta_hall} cannot be used in order to express $\partial_t\bht$ in terms of $\bpi_{\eta}$;  therefore,  it can be interpreted as a constraint between the dynamical variables, which helps us though to express explicitly $\bA_1$ in terms of canonical variables via $\bA_1=\rho^{-1}\bpi_\eta-(\bzt-d_i\bht)\cdot\nb\bA$.  A consistency condition is that this equation holds for all  time, i.e.,  that it is preserved by the dynamics,  
\begin{equation}
[\bpi_\eta-\rho\left(\bA_1+\bzt\cdot\nb \bA-d_i \bht\cdot\nb\bA \right),\Hc_2]=0\,, 
\label{consistency_1}
\end{equation}
where 
\begin{eqnarray}
[f,g]&=&\int d^3x\, \bigg(\frac{\delta f}{\delta\bzt}\cdot\frac{\delta g}{\delta\bpi_\zeta}
-\frac{\delta g}{\delta\bzt}\cdot\frac{\delta f}{\delta\bpi_\zeta}
 \label{ordinary_poisson}\\
&&\hspace{2cm}  
+\frac{\delta f}{\delta\bht}\cdot\frac{\delta g}{\delta\bpi_\eta} 
-\frac{\delta g}{\delta\bht}\cdot\frac{\delta f}{\delta\bpi_\eta}\bigg)\,\nn
\end{eqnarray}
 is the canonical Poisson bracket and 
\begin{eqnarray}
 \Hc_2=\int d^3x \bigg[\frac{1}{2\rho}\big|\bpi_\zeta-\rho\bv\cdot\nb\bzt-\rho\bht\cdot\nb\bA\big|^2
 \nn
\\  -\mathfrak{W}_{hmhd}(\bzt,\bht,\bpi_\eta)\bigg]\,, \label{hamiltonian_hall_perturbed}
\end{eqnarray}  
where $\bA_1$ has been expressed via Eq.~\eqref{pi_eta_hall}.
From \eqref{consistency_1} \eqref{ordinary_poisson} and \eqref{hamiltonian_hall_perturbed} we find
\begin{eqnarray}
-\frac{\partial\mathfrak{W}_{hmhd}}{\partial \bht}&=&d_i\nb\bA\cdot
\bigg\{\frac{\bJ}{\rho}\nb\cdot(\rho\bzt)
\label{consistency_2}\\
&& +\rho\Big[(\bzt-d_i\bht)\cdot\nb\frac{\bJ}{\rho}-\bv\cdot\nb\bht +\bht\cdot\nb\bv 
\nn\\
&&\hspace{1cm}  -\frac{\bJ}{\rho}\cdot\nb\bzt+d_i\frac{\bJ}{\rho}\cdot\nb\bht\Big]\nn\\
&&+\nb\times\nb\times\left[\rho^{-1}\bpi_\eta-(\bzt-d_i\bht)\cdot\nb\bA\right]\bigg\}\,.
\nn
\end{eqnarray} 
%%%%%%%%%%%
Now, let us proceed by computing the Hamiltonian  equations of motion
%%%%%%%
\begin{widetext}
\begin{eqnarray}
\partial_t\bht=&&\frac{\delta\Hc_2}{\delta\bpi_\eta}=-\bv\cdot\nb\bht+\bht\cdot\nb\bv-\frac{\bJ}{\rho}\cdot\nb\bzt+\bzt\cdot\nb\frac{\bJ}{\rho}+d_i\left(\frac{\bJ}{\rho}\cdot\nb\bht-\bht\cdot\nb\frac{\bJ}{\rho}\right)+\frac{\bJ}{\rho^2}\nb\cdot(\rho\bzt)\nn\\
&&+\rho^{-1}\nb\times\nb\times\left[\rho^{-1}\bpi_\eta-(\bzt-d_i\bht)\cdot\nb\bA\right]\,,\label{delta_t_eta}\\
\partial_t\bzt=&&\frac{\delta\Hc_2}{\delta\bpi_\zeta}=\rho^{-1}(\bpi_\zeta-\rho\bv\cdot\nb\bzt-\rho\bht\cdot\nb\bA)\,,\label{delta_t_zeta}\\
\partial_t\bpi_\eta=&&-\frac{\delta\Hc_2}{\delta\bht}=\nb\bA\cdot(\bpi_\zeta-\rho\bv\cdot\nb\bzt-\rho\bht\cdot\nb\bA)+\frac{\partial \mathfrak{W}_{hmhd}}{\partial\bht}\,,\label{delta_t_pi_eta}\\
\partial_t\bpi_\zeta=&&-\frac{\delta\Hc_2}{\delta\bzt}=-\bigg\{\rho\bv\cdot\nb[\rho^{-1}\bpi_\zeta-\bv\cdot\nb\bzt-\bht\cdot\nb\bA]+\rho\bv\cdot\nb(\bht\cdot\nb\bA)+\rho\bv\cdot\nb(\bv\cdot\nb\bzt)+\bJ\cdot\nb\frac{\bpi_\eta}{\rho}\nn\\
&&-\rho\nb\bA\cdot\left[(\bht\cdot\nb\bv)+(\bzt-d_i\bht)\cdot\nb\frac{\bJ}{\rho}+\frac{\bJ}{\rho}\nb\cdot(\rho\bzt)+\nb\times\nb\times(\rho^{-1}\bpi_\eta-\bzt\cdot\nb\bA+d_i\bht\cdot\nb\bA)\right]\nn\\
&&-\rho(\bht\cdot\nb\nb\bA)\cdot\bv-\nb\left[\rho^{-1}\bpi_\eta-(\bzt-d_i\bht)\cdot\nb\bA\right]\cdot\bJ-(\bzt\cdot\nb\nb\bA)\cdot\bJ+d_i(\bht\cdot\nb\nb\bA)\cdot\bJ+\rho\bht\cdot\nb\nb\Phi\nn\\
&&+\nb p\nb\cdot\bzt-\nb\bzt\cdot\nb p-\nb\left(\rho \frac{\partial p}{\partial\rho}\nb\cdot\bzt\right)-d_i\nb p_e \nb\cdot\bht+d_i\nb\bht\cdot\nb p_e+d_i\nb\left( \rho\frac{\partial p_e}{\partial \rho}\nb
\cdot\bht\right)\bigg\}\,. \label{delta_t_pi_zeta}
\end{eqnarray}
\end{widetext}
%%%%%%%%%%%%%%
Combining \eqref{delta_t_eta} with \eqref{pi_eta_hall} and \eqref{dw_hmhd} gives 
%%%%%%%%%%%%%%%%
\begin{eqnarray}
\rho_1=-\nb\cdot(\rho\bzt)\,.
\end{eqnarray}
%%%%%%%%%%%%%%%%
Equation \eqref{delta_t_zeta} is merely the definition of the canonical momentum $\bpi_\zeta$. Exploiting the definitions \eqref{pi_zeta_hall}--\eqref{pi_eta_hall}, the relations \eqref{dv_hmhd} and \eqref{dw_hmhd} and also the stationary momentum equation and Ohm's law, which are given by 
%%%%%
\begin{eqnarray}
\bv\cdot\nb\bv-\rho^{-1}\bJ\times\bB+\rho^{-1}\nb p=0\,,\\
-\nb\Phi+\left(\bv-d_i\frac{\bJ}{\rho}\right)\times\bB+\rho^{-1}\nb p_e=0\,,
\end{eqnarray}
%%%%%
we can corroborate that \eqref{delta_t_pi_eta} and \eqref{delta_t_pi_zeta} give the perturbed Ohm's law and momentum equation, respectively. %Therefore, the Hamiltonian \eqref{hamiltonian_hall_perturbed} describes correctly the linearized HMHD dynamics and, consequently,  the condition $-\int d^3x\,\mathfrak{W}_{hmhd}\geq0$ suffices for stability. 
Note that $\mathfrak{W}_{hmhd}$ is not yet fully expressed in terms of the displacement vectors $\bzt$ and $\bht$ due to $\bpi_\eta$,  which appears explicitly in its expression. We can overcome this by combining the consistency condition  \eqref{consistency_2} with the Hamiltonian  equations \eqref{delta_t_eta} and \eqref{delta_t_pi_eta} to find
%%%%%%%%%%%%%%%%
\begin{eqnarray}
\partial_t\bA_1=\partial_t(\bzt-d_i\bht)\times\bB_0\,.
\end{eqnarray}
Integrating in time would in general introduce a stationary vector, however this should vanish because otherwise time independent terms would appear in the linearized dynamical equations. Therefore $\bA_1=(\bzt-d_i\bht)\times\bB_0$ or $\bB_1=\nabla\times[(\bzt-d_i\bht)\times\bB_0]$,  which is the well-known solution of the perturbed induction equation (see \cite{Hirota2006}). This expression is similar with the corresponding expression in ideal MHD. The difference is the appearance of the displacement vector $\bht$ multiplied by $d_i$; so, the MHD result can be recovered in the limit $d_i\rightarrow 0$. This is an anticipated result, since the fluid velocity in the MHD induction equation is replaced by $\bv-d_i\bJ$ in the HMHD case. Finally, since \eqref{hamiltonian_hall_perturbed} describes correctly the dynamics, we conclude that 
%%%%%
\begin{eqnarray}
-\int d^3x\,\mathfrak{W}_{hmhd}(\bzt,\bht)\geq 0\,, 
\end{eqnarray}
%%%%%
where $\mathfrak{W}_{hmhd}(\bzt,\bht)$ is given by \eqref{W_hmhd} with $\bA_1=(\bzt-d_i\bht)\times\bB_0$, is sufficient for stability. Note that the term containing $\nb\Phi_1$ can be neglected in view of $\nb\cdot(\rho\bht)=0$ and $\bht\cdot \hat{\mathrm{n}}\big|_{\partial D}=0$.

%%%%%%%%%%%%%%%%%%%%%%%%%%%%%
%%%%%%%%%%%%%%%%%%%%%%%%%%%%%
%%%%%%%%%%%%%%%%%%%%%%%%%%%%%

\section{Conclusions}

In this paper, we derived sufficient stability criteria, exploiting the Hamiltonian structure of the XMHD model. The energy-Casimir,  dynamically accessible, and Lagrangian methods were used. Using the EC method we ascertained that indefinite terms appear in the second variation of the EC functional occurring due the vorticity-magnetic field coupling induced by the form of the Casimir invariants. We side-stepped  this problem by  considering equilibria with purely toroidal flow or special perturbations, assumptions that enable the removal of the indefiniteness. To study stability under three-dimensional perturbations we employed the DA method, which  allows the  study of  stability of generic equilibria by  restricting  the perturbations to be tangent on the Casimir leaves. Such perturbations are consistent with the physics under consideration.  Finally we developed a Lagrangian stability analysis of the quasi-neutral two-fluid model written in MHD-like variables, namely the Lagrangian counterparts of the center of mass velocity and current density. Subsequently employing the Lagrange-Euler map we jumped to the Eulerian viewpoint and upon performing a Legendre transformation we found the Hamiltonian of the linear dynamics. Considering massless electrons, the definition of one of the two canonical momenta led to a relation between the perturbed magnetic potential and canonical variables. Requiring this relation to be preserved by the dynamics gave rise to a dynamical constraint;  whence we found the solution to the perturbed induction equation, namely,  $\bB_1=\nabla\times[(\bzt-d_i\bht)\times\bB]$. In addition,  we generalized the HMHD energy principle so as to include the electron pressure contribution.

%%%%%%%%%%%%%%%%%%%%%%%%%%%%%
%%%%%%%%%%%%%%%%%%%%%%%%%%%%%
%%%%%%%%%%%%%%%%%%%%%%%%%%%%%

\section*{Acknowledgements}

This work has been carried out within the framework of the EUROfusion Consortium and has received funding from the Euratom research and training programme 2014-2018 and 2019-2020 under grant agreement No 633053 as well as from the National Programme for the Controlled Thermonuclear Fusion,
Hellenic Republic. The views and opinions expressed herein do not necessarily reflect those of the European Commission. D.A.K.\  was financially supported by the General Secretariat for Research and Technology (GSRT) and the Hellenic Foundation for Research and Innovation (HFRI). P.J.M.\  was supported by the U.S. Department of Energy under Contract No.\  DE-FG05-80ET-53088. The authors warmly acknowledge the hospitality of the Numerical Plasma Physics Division of Max Planck IPP,  Garching, where a portion of this research was done.

\medskip
\begin{appendices}

%%%%%%%%%%%%%%%%%%%%%%%%%%%%%
%%%%%%%%%%%%%%%%%%%%%%%%%%%%%
%%%%%%%%%%%%%%%%%%%%%%%%%%%%%
\begin{widetext}
\section{Intermediate result for $\delta^2 \Hc_{da}$}
\label{appendix_A}
Inserting expressions \eqref{d^2rho_da}--\eqref{d^2B_da} into \eqref{d2H_da}, we readily find
%%%%%%%%%%
\begin{eqnarray}
&&\delta^2\Hc_{da}=\int_Vd^3x\,\bigg\{\rho\big|-\nb \mathrm{g}_0+\bzt\times\bom+\bht\times\bB^*-\frac{\bv}{\rho} \nb\cdot(\rho\bzt)\big|^2+|\delta \bB_{da}|^2+d_e^2\rho^{-1}|\nb\times\delta\bB_{da}|^2\nn\\
&&+\rho^{-1}\left(c_s^2-|\bv|^2-d_e^2\frac{|\bJ|^2}{\rho^2}\right)[\nb\cdot(\rho\bzt)]^2-\bzt\cdot\nb\left(h+\frac{|\bv|^2}{2}+d_e^2\frac{|\bJ|^2}{2\rho^2}\right)\nb\cdot(\rho\bzt)-\bht\cdot(\bv\times\bB^*)\nb\cdot(\rho\bzt)\nn\\
%+2\bzt\cdot(\bv\times\bom)\nb\cdot(\rho\bzt)\nn\\
&&-\rho^{-1}\bht\cdot(-d_i\bJ\times\bB^*+d_e^2\bJ\times\bom)\nb\cdot(\rho\bzt)
-\rho(\bzt\times\bv)\cdot\nb\times\left(\bzt\times\bom\right)-\rho(\bzt\times\bv)\cdot\nb\times\left(\bht\times\bB^*\right)\nn\\
&&-\rho(\bht\times\bv)\cdot\nb\times\left[(\bzt-d_i\bht)\times\bB^*\right]-\left[(\bzt-d_i\bht)\times\bJ\right]\cdot\nb\times\left[(\bzt-d_i\bht)\times\bB^*\right]-d_e^2\rho(\bht\times\bv)\cdot\nb\times\left(\bht\times\bom\right)\nn\\
&&-d_e^2[(\bzt-d_i\bht)\times\bJ]\cdot\nb\times\left(\bht\times\bom\right)-d_e^2(\bht\times\bJ)\cdot\nb\times\left(\bzt\times\bom\right)-d_e^2(\bht\times\bJ)\cdot\nb\times\left(\bht\times\bB^*\right)\bigg\}\,. \label{d2H_da_xmhd}
\end{eqnarray}
%\end{widetext}
As a simple application let us consider a stationary  axisymmetric equilibrium with purely toroidal flow and variations with perturbation vectors that never leave the surfaces $\psi^*=const.$, i.e. $\bzt\cdot\nb\psi^*=0$ and $\bht\cdot\nb\psi^*=0$. To find the equilibrium conditions we set  $\partial_t\rightarrow 0$ and $\bv=rv_\phi\nb\phi$ in  \eqref{mom_eq}--\eqref{ind_eq}. Then the XMHD equations reduce to
%\begin{widetext}
\begin{eqnarray}
&&r^{-1}v_\phi\nb(r v_\phi)-\nb\left(h+\frac{|\bv|^2}{2}+d_e^2\frac{|\bJ|^2}{2\rho^2}\right)
-\rho^{-1}\left[\frac{\Delta^*\psi}{r^2}\nb\psi^*+\frac{B_\phi^*}{r}\nb(rB_\phi)-\nb(rB_\phi)\cdot(\nb\psi^*\times\nb\phi)\nb\phi\right]=0 \,,\label{stationary_mom}\\
&&
r^{-1}v_\phi\nb\psi^*-\rho^{-1}\bigg\{d_i\left[-\frac{\Delta^*\psi}{r^2}\nb\psi^*-\frac{B_\phi^*}{r}\nb(rB_\phi)+\nb(rB_\phi)\cdot(\nb\psi^*\times\nb\phi)\nb\phi\right]\nn\\
&&\hspace{7cm}  +d_e^2\left[\frac{\Delta^*\psi}{r^2}\nb(r v_\phi)-\nb(rB_\phi)\cdot(\nb(rv_\phi)\times\nb\phi)\nb\phi\right]\bigg\}=\nb\tilde{\Phi}\,, \label{stationary_ind}
\end{eqnarray}
\end{widetext}
where $\tilde{\Phi}=\Phi-d_ih_e+d_e^2\rho^{-1}\bv\cdot\bJ-d_id_e^2\rho^{-2}|\bJ|^2$, with $\Phi$ and $h_e$ being the equilibrium electrostatic potential and electron specific enthalpy, respectively.
Projecting Eq.~\eqref{stationary_mom} and Eq. \eqref{stationary_ind} along $\nb\phi$ we find 
\begin{eqnarray}
\nb(rB_\phi)\cdot(\nb\psi^*\times\nb\phi)=0\,,\Leftrightarrow rB_\phi=F(\psi^*)\,. \label{rB_phi}
\end{eqnarray}
Similarly projecting Eq. \eqref{stationary_ind} and using result \eqref{rB_phi} we find 
\begin{eqnarray}
\nb(rv_\phi)\cdot(\nb\psi^*\times\nb\phi)=0\,,\Leftrightarrow rv_\phi=G(\psi^*)\,. \label{rv_phi}
\end{eqnarray}
Equations \eqref{rB_phi} and \eqref{rv_phi} imply   $\bJ=-\Delta^*\psi\nb\phi+F'(\psi^*)\nb\psi^*\times\nb\phi$ and $\bom=G'(\psi^*)\nb\psi^*\times\nb\phi$, respectively. Therefore, $\bJ\cdot\nb\psi^*=\bom\cdot\nb\psi^*=0$. This means that all three vector fields $\bv$, $\bB^*$ and $\bJ$ lie on common flux surfaces labeled by $\psi^*$. This property of common flux surfaces was crucial for the derivation of a sufficient stability criterion in the context of MHD \cite{Throumoulopoulos2007} for a three-dimensional incompressible displacement vector field. It is thus interesting to pursue the investigation of this possibility also in the context of XMHD in the future. As regards the current application,  we  confine the perturbation vectors to be tangent to the characteristic surfaces. Also note that using the result \eqref{rB_phi} and projecting \eqref{stationary_mom} along $\bB^*$ we find 
\begin{eqnarray}
\nb\tilde{h}\cdot(\nb\psi^*\times\nb\phi)=0\,, \Leftrightarrow \tilde{h}=\tilde{h}(\psi^*)\,.\label{h}
\end{eqnarray}
%In view of \eqref{rB_phi}, \eqref{rv_phi} and \eqref{h} the momentum balance equation leads to
%\begin{eqnarray}
%\frac{GG'(\psi^*)}{r^2}-\tilde{h}'(\psi^*)-\frac{\Delta^*\psi}{\rho r^2}-\frac{F'(\psi^*)B_\phi^*}{r\rho}=0\,.\label{stationary_mom_2}
%\end{eqnarray}
%From the definition of $\bB^*$ (Eq. \eqref{B*}) we can find 
%\begin{eqnarray}
%B_\phi^*=r^{-1}F(\psi^*)-d_e^2 r\nb\left[\frac{F'(\psi^*)}{\rho r^2}\nb\psi^*\right]\,. \label{B_phi*}
%\end{eqnarray}
%Inserting \eqref{B_phi*} into \eqref{stationary_mom_2} we find a Grad-Shafranov equation
%\begin{eqnarray}
%&&d_e^2r^2F'\nb\left(\frac{F'}{\rho r^2}\nb\psi^*\right)=FF'+\rho r^2 \tilde{h}'(\psi^*)+\Delta^*\psi-\rho GG'\,,\label{GS}
%\end{eqnarray}
%coupled to 
%\begin{eqnarray}
%\Delta^*\psi=\frac{\rho}{d_e^2}(\psi-\psi^*)\,, \label{GS_psi}
%\end{eqnarray}
%which follows from \eqref{B*} and 
%\begin{eqnarray}
%h(\rho)=\tilde{h}(\psi^*)-\frac{v_\phi^2}{2}-d_e^2\frac{|\bJ|^2}{2\rho^2}\,. \label{Bernoulli_2}
%\end{eqnarray}
For equilibria with purely toroidal flows, subject to perturbations with displacement vectors tangent to the common surfaces,  it easy to understand that every product of the form $\mathbf{b}_i\times\mathbf{c}_j$ where $\mathbf{b}=(\bzt,\bht)$ and $\mathbf{c}=(\bv,\bB^*,\bJ)$, will be parallel to the vector $\nb\psi^*$ at each surface point, i.e.\  $\mathbf{b}_i\times\mathbf{c}_j=g_{ij}(r,z)\nb\psi^*$. Therefore every vector of the form $\nb\times(\mathbf{b}\times\mathbf{c})$ will be $\nb g\times\nb\psi^*$ and consequently every term  $(\mathbf{b}_i\times\mathbf{c}_j)\cdot\nb\times(\mathbf{b}_k\times\mathbf{c}_\ell)$ in \eqref{d2H_da_xmhd} will vanish. The same holds also for terms of the form $\mathbf{b}_i\cdot(\mathbf{c}_j\times\mathbf{c}_k)$, since $(\mathbf{c}_j\times\mathbf{c}_k)$ is normal to the characteristic surfaces at each point, if not zero. In addition the term containing $\bzt\cdot\nb\tilde{h}$ will vanish as well due to \eqref{h}. A rigorous proof can be carried out upon writing 
%\begin{eqnarray}
$\bzt=r\zeta_\phi\nb\phi+(\bzt\cdot\bB^*_p)/|\bB_p^*|^2\bB_p^*\,,$
%\end{eqnarray}
which is a general representation of vectors tangent to $\psi^*=const.$ surfaces, and similarly for $\bht$; then computing every single term in \eqref{d2H_da_xmhd}, leading eventually to %\eqref{stab_par_pert}. 
\begin{eqnarray}
\delta^2\Hc_{da}=\int_Vd^3x\,\bigg\{\rho\big|\delta\bv_{da}+\frac{rv_\phi}{\rho}\delta\rho_{da}\nb\phi\big|^2\nn\\
+|\delta \bB_{da}|^2
+d_e^2\rho^{-1}|\nb\times\delta\bB_{da}|^2\nn
\\
+\rho^{-1}\left(c_s^2-v_\phi^2-d_e^2\frac{|\bJ|^2}{\rho^2}\right)[\nb\cdot(\rho\bzt)]^2\bigg\}\,.\label{stab_par_pert}
\end{eqnarray}
As a result, $c_s^2-v_\phi^2-d_e^2{|\bJ|^2}/{\rho^2}>0$ is sufficient for stability and also for the ellipticity of the equilibrium Grad-Shafranov-Bernoulli equations. Actually for the ellipticity of the equilibrium system, the condition $c_s^2-d_e^2{|\bJ_p|^2}/{\rho^2}>0$ is sufficient, as was  shown in \cite{Kaltsas2019}. \\ \indent
\section{Expansion of the internal energy}
\label{appendix_B}
The difficulty in this expansion is that the Jacobians contain a dependence on the gradients of the fluid trajectories; therefore,  we need to know how to differentiate the $\Jc$'s, because the expansion of the internal energy is effected through the expansion
%%%%%%%%
\begin{eqnarray}
\Jc_s=\Jc_{s0}+\frac{\partial\Jc_s}{\partial q^i_{s,j}}\frac{\partial\zeta_s ^i}{\partial a_s^j}+\frac{1}{2}\frac{\partial^2 \Jc_s}{\partial q_{s,k}^i\partial q_{s,\ell}^j}\frac{\partial\zeta_s^i}{\partial a_s^k}\frac{\partial\zeta_s^j}{\partial a_s^\ell}\,,
\end{eqnarray}
%%%%%%%%%%
where $q_{s,j}^i:=\frac{\partial q_s^i}{\partial a_s^j}$. The derivatives of the Jacobian are $\frac{\partial \Jc_s}{\partial q_{s,j}^i}=C_{si}^{\;\,j}$, where $C_{si}^{\;\, j}=\frac{1}{2}\epsilon_{i\ell k}\epsilon^{jmn}\frac{\partial q_{s}^\ell}{\partial a_s^m}\frac{\partial q_{s}^k}{\partial a_s^n}$ are the cofactors of $\partial q_s^i/\partial a_s^j$ in $\Jc_s$. Following the procedure in \cite{Morrison1998} and \cite{Newcomb1962}, we find 
%%%%
\begin{eqnarray}
\hspace{-6mm}\Jc_{s1}=\Jc_{s0}\frac{\partial \zeta_s^i}{\partial q_{s}^i}\,,\; \Jc_{s2}=\frac{\Jc_{s0}}{2}\left[\left(\frac{\partial \zeta_{s}^i}{\partial q_s^i}\right)^2-\frac{\partial \zeta_{s}^i}{\partial q_s^j}\frac{\partial \zeta_{s}^j}{\partial q_s^i}\right]\,.
\end{eqnarray}
%%%%%%%%%%%%%%%
With these expressions at hand,  we can find the second order perturbation of the internal energies in terms of the displacement vectors to be 
%%%%%%%%%
%
\begin{align}
U_{s2}&=\frac{n_{s0}}{2\Jc_{s0}}\bigg\{\frac{\partial U_s}{\partial n_s}\bigg[\left(\frac{\partial \zeta^i}{\partial q_s^i}+\alpha_s \frac{\partial \eta^i}{\partial q_s^i}\right)^2\nn\\
&+\left(\frac{\partial \zeta^i}{\partial q_s^j}+\alpha_s \frac{\partial \eta^i}{\partial q_s^j}\right)\left(\frac{\partial \zeta^j}{\partial q_s^i}+\alpha_s \frac{\partial \eta^j}{\partial q_s^i}\right)\bigg]\nn\\
& +\frac{n_{s0}}{\Jc_{s0}}\frac{\partial^2 U_s}{\partial n_s^2}\left(\frac{\partial \zeta^i}{\partial q_s^i}+\alpha_s \frac{\partial \eta^i}{\partial q_s^i}\right)^2\bigg\}\,. \label{sec_order_internal_energy}
\end{align}

\begin{widetext}
\section{Eulerian displacement vectors}
\label{appendix_C}

 Let us begin with the Lagrange-Euler map and its inverse in order to understand how $\bQ$, $\bD$,  and the displacements $\bzt$, $\bht$ are mapped into  the Eulerian coordinates. From   \eqref{L-E_map_1} and \eqref{material_xmhd_vars_1}--\eqref{inverse_material_xmhd_vars} we can effectively construct every map we need. For  example,
%%%
\begin{eqnarray}
\dot{\bQ}(\ba_i,\ba_e,t)&=&\frac{m_i}{m}\left(\bv+\frac{m_e}{men}\bJ\right)\bigg|_{\bx=\bq_{i}(\ba_i,t)}+\frac{m_e}{m}\left(\bv-\frac{m_i}{men}\bJ\right)\bigg|_{\bx=\bq_{e}(\ba_e,t)}\,,\nn\\
\dot{\bD}(\ba_i,\ba_e,t)&=&\left(\bv+\frac{m_e}{men}\bJ\right)\bigg|_{\bx=\bq_{i}(\ba_i,t)}-\left(\bv-\frac{m_i}{men}\bJ\right)\bigg|_{\bx=\bq_{e}(\ba_e,t)}\,. \label{inverse_dot_Q_D}
\end{eqnarray}
%%%
%For the computation of the equilibrium $\dot{\bQ}_0$ and  $\dot{\bD}_0$ we evaluate at $\bq_{i0}$ and  $\bq_{e0}$,  while the Eulerian quantities $n, \bv,\bJ$ correspond to an Eulerian equilibrium. 
If these expressions are computed at $\ba_e=\bq_e^{-1}(\bq_i(\ba_i,t),t)$ as in the Lagrangian \eqref{material_xmhd_Lagrangian}, then at equilibrium we have $\dot{\bQ}_0(\ba_i,t)=\bv(\bx)\big|_{\bx=\bq_{i0}(\ba_i,t)}$ and $\dot{\bD}_0(\ba_i,t)=e^{-1}n^{-1}(\bx)\bJ(\bx)\big|_{\bx=\bq_{i0}(\ba_i,t)}$. For the Eulerianization of the displacement vectors, we define their Eulerian displacements $\tbzt$ and  $\tbht$ by 
%%%%%%%%%
\begin{eqnarray}
\bzt(\ba_i,\ba_e,t)&=&\frac{m_i}{m}\left[\tbzt(\bx,t)+\frac{m_e}{m}\tbht(\bx,t)\right]_{\bx=\bq_{i0}(\ba_i,t)}+\frac{m_e}{m}\left[\tbzt(\bx,t)-\frac{m_i}{m}\tbht(\bx,t)\right]_{\bx=\bq_{e0}(\ba_e,t)}\,,\nn\\
\bht(\ba_i,\ba_e,t)&=&\left[\tbzt(\bx,t)+\frac{m_e}{m}\tbht(\bx,t)\right]_{\bx=\bq_{i0}(\ba_i,t)}-\left[\tbzt(\bx,t)-\frac{m_i}{m}\tbht(\bx,t)\right]_{\bx=\bq_{e0}(\ba_e,t)}\,.\label{euler_displacents}
\end{eqnarray}
Taking the time derivatives of \eqref{euler_displacents} with $\ba_i$ and $\ba_e$ held constant, we find 
\begin{eqnarray}
\dot{\bzt}(\ba_i,\ba_e,t)&=&\partial_t\tbzt(\bx,t)+\bv\cdot\nb\tbzt(\bx,t)+\frac{m_im_e}{m^2}\bw\cdot\nb\tbht(\bx,t)\,,\nn\\
\dot{\bht}(\ba_i,\ba_e,t)&=&\partial_t\tbht(\bx,t)+\bv\cdot\nb\tbht(\bx,t)+\bw\cdot\nb\tbzt(\bx,t)+\frac{m_e^2-m_i^2}{m^2}\bw\cdot\nb\tbht(\bx,t)\,,\label{time_der_euler_displacents}
\end{eqnarray}
where $\bw:=\bJ/(en)$ and we have made use of $\bv_s(\bx)=\bv(\mathbf{x})+\alpha_s\bJ(\mathbf{x})/(en(\mathbf{x}))=\dot{\bq}_{s0}(\ba_s,t)\big|_{\ba_s=\bq_{s0}^{-1}(\mathbf{x},t)}$. This result, along with \eqref{inverse_dot_Q_D}, is used for the derivation of \eqref{two-fluid_perturbed_Lagrangian}. Taking the first variation of \eqref{inverse_dot_Q_D} and identifying 
\[
\delta\dot{\bQ}=\dot{\bzt}\,, \qquad \delta\dot{\bD}=\dot{\bht}\,, \qquad 
\mathrm{and}\qquad 
\delta \bq_s(\ba_s,t)\big|_{\ba_s=\bq_s^{-1}(\bx,t)}=\tbzt+\alpha_s\tbht\,,
\]
 after some manipulations we find 
\begin{eqnarray}
\dot{\bzt}=\delta \bv+\tbzt\cdot\nb\bv+\frac{m_im_e}{m^2}\tbht\cdot\nb\bw\,,\quad
\dot{\bht}=\delta \bw+\tbht\cdot\nb\bv+\tbzt\cdot\nb\bw+\frac{m_e^2-m_i^2}{m^2}\tbht\cdot\nb\bw\,. \label{var_dot_Q-D}
\end{eqnarray}
Combining \eqref{var_dot_Q-D} with \eqref{time_der_euler_displacents} we arrive at \eqref{delta_v} and \eqref{delta_w}.
\end{widetext}
\end{appendices}
\medskip

%%%%%%%%%%%%%%%%%%%%%%%%%%%%%
%%%%%%%%%%%%%%%%%%%%%%%%%%%%%
%%%%%%%%%%%%%%%%%%%%%%%%%%%%%

\end{document}